\documentclass[prx,floatfix,footinbib,notitlepage,onecolumn,superscriptaddress,longbibliography]{revtex4-2}
\usepackage[makeroom]{cancel}
\usepackage[euler]{textgreek}
\usepackage[bbgreekl]{mathbbol}
\usepackage{graphicx}
\usepackage[explicit]{titlesec}
\usepackage[utf8]{inputenc}
\usepackage{amsmath,amssymb,mathtools}
\usepackage{bbold}
\usepackage{graphicx}
\usepackage{bm,color}
\usepackage{natbib,hyperref}
\usepackage[capitalise]{cleveref}
\usepackage{enumitem}
\usepackage{adjustbox}
\usepackage{mathrsfs}
\usepackage[normalem]{ulem}
\usepackage{xspace}
\usepackage[dvipsnames]{xcolor}
\usepackage[displaymath, mathlines]{lineno}

\hypersetup{breaklinks=true,colorlinks=true,linkcolor=blue,urlcolor=blue,citecolor=blue}

\newcommand{\ket}[1]{\left| #1 \right\rangle}
\newcommand{\bra}[1]{\left\langle #1 \right|}
\DeclarePairedDelimiterX\braket[2]{\langle}{\rangle}{#1 \delimsize\vert #2}
\newcommand{\abs}[1]{\left| #1 \right|}

\newcommand{\norm}[1]{\ensuremath{\left| \left|#1 \right| \right|}}

\def\O{{\cal{O}}}
\def\t{\theta}
\def\X{\bar{X}}
\def\P{\bar{P}}
\def\n{\bar{n}}
\def\Q{\bar{Q}}
\def\S{\bar{S}}
\def\hf{\hat{f}}
\def\hg{\hat{g}}
\def\hh{\hat{h}}
\def\hphi{ \hat{\phi} }
\def\F{{\cal{F}}}
\def\T{{\cal{T}}}
\def\e{\epsilon}
\def\tF{\bar{\cal{F}}}

\def\gZZ{~\mathclap{Z}Z}
\def\gZZZ{~Z\mathclap{Z}Z}

\definecolor{vermillion}{rgb}{0.86, 0.18, 0.01}

\begin{document}

\title{Qumode transfer between continuous and discrete variable devices}
\author{Alexandru Macridin}
\affiliation{Fermi National Accelerator Laboratory, Batavia, IL, 60510, USA}
\author{Andy C.~Y.~Li}
\affiliation{Fermi National Accelerator Laboratory, Batavia, IL, 60510, USA}
\author{Panagiotis Spentzouris}
\affiliation{Fermi National Accelerator Laboratory, Batavia, IL, 60510, USA}
\date{\today}

\begin{abstract}

Transferring quantum information between different types of quantum hardware is crucial for integrated quantum technology. In particular, converting information between continuous-variable (CV) and discrete-variable (DV) devices enables many applications in quantum networking, quantum sensing, quantum machine learning, and quantum computing.
This paper addresses the transfer of CV-encoded information between CV and DV devices.
We present a resource-efficient method for encoding CV states and implementing CV gates on DV devices, as well as two measurement-based protocols for transferring CV states between CV and DV devices. The success probability of the transfer protocols depends on the measurement outcome and can be increased to near-deterministic values by adding  ancillary qubits to the DV devices.

\end{abstract}

\maketitle

\section{Introduction}

In the past decade, there has been a major focus on developing quantum technology that holds immense potential for revolutionizing communication, sensing, and computing domains. A wide variety of platforms, including superconducting circuits, microwave cavities, optical systems, trapped ions, atoms, spins, and others, have been used to process quantum information~\cite{Ladd_Nature_2010, kurizki_2015}. In these systems, information is encoded in a set of quantum states that can be discrete (qubits and qudits) or continuous (qumodes). Depending on the specific area of application, both types of encoding have their advantages and disadvantages. For the development of integrated quantum technology, it is essential to have the capability to transfer information between all types of quantum devices.
While effort has previously been devoted to processing logical qubits encoded on continuous-variable (CV) devices~\cite{Cochrane_PRA_1999,Jeong_PRA_2002,Ralph_PRA_2003,Vlastakis2013}, we consider an alternative perspective here: encoding and processing continuous-variable information on discrete-variable (DV) devices. We present a method to encode CV information into DV devices, along with two measurement-based transfer protocols to convert information between CV and DV devices.

The ability to develop hybrid DV-CV technology and convert encoded information between platforms is crucial for building complex systems, such as the quantum internet~\cite{Pirandola_Nat_2016} and quantum sensor networks~\cite{Zhang_2021}. For instance, while superconducting chips are better for data processing, optical devices are currently best for long-distance communication and are easily scalable. Various hybrid DV-CV methods have recently been proposed for quantum teleportation~\cite{Takeda_Nat_2013, Andersen_PRL_2013, Andersen_NP_2015}, entanglement distillation~\cite{brask_PRL_2010}, and quantum computing~\cite{Zhang_PRA_2019, Gan_prl_2020}. In addition, methods to encode qubits in CV devices, such as cat states~\cite{Cochrane_PRA_1999,Jeong_PRA_2002,Ralph_PRA_2003,Vlastakis2013}, and GKP states~\cite{Gottesman_PRA_2001}, have also been proposed to increase qubit resilience to errors. Furthermore, significant effort has been put into developing methods to entangle DV and CV qubits \cite{Jeong_NP_2014, Morin_NP_2014, Huang_NJP_2019} and to convert DV and CV qubits from one to the other via teleportation protocols~\cite{Ulanov_PRL_2017, Darras_NatPh_2023, Jeong_NatPh_2023}.

Aside from the possibility of encoding qubits, the CV devices have the ability to  process information encoded in the continuous bases formed by the eigenvectors of the field quadrature operators, known also as {\em{qumode}} encoding. CV quantum computing~\cite{Braunstein_RMP_2005,Weedbrook_RMP84_2012} is universal~\cite{Lloyd_PRL82_1999}, meaning that any unitary transformation  generated by a polynomial function of the quadrature operators can be decomposed into a finite number of gates drawn from a finite set of gates. Recent advancements in photonic chips~\cite{Arrazola_Nat_2021,Madsen_Nature_2022}  and the availability of CV quantum software, such as Strawberry Fields~\cite{Killoran_Quantum_2019}, indicate that this is an active and rapidly evolving research area. A significant amount of effort has been devoted to the development of quantum CV algorithms. Currently, the CV algorithms  address a wide range of problems, such as scalar field simulations~\cite{marshall_pra_2015,yeter_PRA_2022}, spin simulations~\cite{Zhang_prl_2021},
attractive Bose-Hubbard simulations~\cite{Yalouz_Quantum_2021}, partial differential equations~\cite{Arrazola_pra_2019}, quantum approximate optimization algorithm~\cite{Verdon2019AQA}, Grover's search~\cite{pati2000quantum} and the Deutsch-Jozsa problem~\cite{Adcock_2009}.
There is also growing interest in employing CV systems in quantum machine learning (QML) methods~\cite{Lau_PRL_2017,Killoran_PRR_2019}.

In this paper, we address the encoding of qumodes in DV devices and the conversion of qumodes between CV and DV devices. A qumode is a quantum state expressed in an infinite basis set. Therefore, transferring qumodes to a finite qubit device is generally an ill-posed problem. However, for most practical purposes, we can impose a boson occupation cutoff, $N_b$, such that the contribution of states with more than $N_b$ bosons is negligible. One simple way to encode such a truncated qumode to a DV device is by mapping the boson number states with $n<N_b$ to the DV computational basis states. However, this direct encoding may have limited usefulness because information encoded in this way cannot be easily processed on the DV device. This is because information encoded in qumodes is generally processed by employing gates that are functions of the quadrature operators, and these gates have a dense matrix representation in the Fock basis. Consequently, implementing CV gates on a DV device requires a lengthy decomposition (on the order of
$\O(4^{n_q})$, where $n_q$ is the size of DV register) into elementary single-qubit and two-qubit gates~\cite{Barenco_pra_1995,shende_pra69_2004,krol_2021_arxiv,macridin_pra_2021, Sawaya2020}. To achieve effective encoding of qumodes onto DV devices, we not only need to map the qumode's state onto a DV device but also to efficiently implement CV gates on DV devices (i.e., by employing only a small number of elementary gates).

Continuous variable computation can be achieved using purely DV systems when the  information to be processed can be encoded in states that can be truncated with controlled accuracy in boson number basis during evolution~\cite{footnote_cvdv1}.
To encode qumodes on DV devices, we take advantage of the properties of their wavefunction at large argument and use the Nyquist-Shannon expansion of functions with support on finite intervals~\cite{Shannon_1949} to represent them in a discrete quadrature basis. We will call the qumodes mapped onto DV systems in this way {\em{discrete qumodes}}. This encoding has high accuracy~\cite{macridin_pra_2021}, and allows for a straightforward, polynomial scaling implementation of CV gates on DV devices.

We present two transfer protocols: one for transferring CV qumodes to their corresponding discrete representation on DV devices, and another for transferring discrete DV qumodes to CV devices.
Both protocols are modifications of the \emph{one-qubit} CV teleportation protocol described in~\cite{Menicucci_PRL97_2006,Weedbrook_RMP84_2012}. They involve entangling the two systems, measuring the first system, and manipulating the second system using operations that depend on the measurement outcome, as shown diagrammatically in~\cref{fig:diagcvdv,fig:diagdvcv}.
The transfer protocols are non-deterministic and require post-selection since the  probability of success, defined in~\cref{sec:teleprotocols}, depends on the measurement outcome and is smaller than one.
However, the probability of success can be increased by using an ancillary DV register. We call our protocols \emph{near-deterministic} because the probability of success can be brought exponentially
close to one by increasing the number of ancilla qubits.
For example, we find that a CV state with a boson number cutoff $N_b=100$ can be transferred with an accuracy of $\O(10^{-7})$ on a DV register of $8$ qubits with a success probability of $0.99$ ($0.999$) using an ancillary register of $13$ ($20$) qubits. After transfer, the ancillary register can be discarded. Furthermore, the transfer protocols presented here might find immediate or near-future applications when used in the non-deterministic regime for qumodes with cutoff $N_b<20$, since in this case, the total number of required DV  qubits is $\sim 4 - 6$.

The transfer protocols introduced here hold the potential to facilitate the development of CV-DV hybrid hardware for processing CV-encoded information. We believe that our method for converting qumodes between CV and DV devices offers a broad spectrum of potential applications. For instance, quantum sensor networks could benefit from processing data collected by sensors with CV encoding on superconducting QPUs. Qubit based quantum machine learning algorithms  can increase their expressivity by including continuous-variable data encoding. The quantum tomography of CV states~\cite{Lvovsky_RMP_2009} can be reduced  to an equivalent qubit system tomography problem. Transferring DV states to CV registers opens  possibilities for non-Gaussian state preparation and the implementation of non-Gaussian operations on CV platforms. Unique measurement-based quantum algorithms~\cite{Raussendorf_PRL_2001,Zhang_PRA_2006,Gu_PRA_2009,Menicucci_PRA_2011} that use hybrid CD-DV cluster states might be developed, which could be particularly useful for simulating field theories. Nevertheless, further investigation is needed to assess these potential applications.

This paper is organized as follows: In \cref{sec:cv}, we define the qumode and briefly introduce the gates required for CV quantum computing. In \cref{sec:cvdvrep}, we introduce the discrete representation of qumodes on qubit devices. In \cref{sec:teleprotocols}, we present the protocols that transfer qumodes between CV and DV devices. In \cref{sec:ancillas}, we show how to use an ancillary qubit register to increase the success probability of the transfer protocols. Finally, in \cref{sec:conclusions}, we present a summary of our results and the conclusions.

\section{CV states and CV quantum computing}
\label{sec:cv}

Qumodes are vectors belonging to the Hilbert space of square integrable functions, $L^2(\mathbb{R})$.
The observables associated with qumodes are generated by the quadrature operators. We denote the quadrature operators
by $X$ and $P$ because they are equivalent to
the position and the momentum operators of a harmonic oscillator, respectively, obeying  the canonical commutation relations $\left[X,P\right]=i$.
The eigenvectors $\{\ket{x}\}$ of $X$ ($X\ket{x}=x\ket{x}$) and the eigenvectors
$\{\ket{p}\}$ of $P$ ($P\ket{p}=p\ket{p}$), constitutes  continuous basis sets and are connected by the  Fourier transform
\begin{align}
\label{eq:ftdef1}
\ket{p}&=\frac{1}{\sqrt{2 \pi}}\int_{-\infty}^{\infty}    d x e^{i p x} \ket{x}.
\end{align}

Aside form  continuous basis sets, $L^2(\mathbb{R})$ also admits denumerable bases,
like the ones formed by boson number states, also known as Fock states.  The Fock states
are eigenvectors of the harmonic oscillator Hamiltonian and of the boson number operator $a^{\dagger}a$,
where
\begin{align}
\label{eq:aop}
a=\left(\sqrt{\mu}X+iP/\sqrt{\mu}\right)/\sqrt{2},
\end{align}
and $\mu$ is the boson mass.
For example, in optical  devices the bosons are the photons, while in other
platforms, like trapped ion devices, the bosons can be the vibrational modes (phonons)~\cite{Chen_ChiesePhB_2021}.

CV computation employs operators with continuous spectra
to process the data encoded in qumode states.  It has been shown~\cite{Lloyd_PRL82_1999} that the evolution of any Hamiltonian that is a polynomial function of $X$ and $P$ can be simulated using only a small number of gate types.
For example, a sufficient set of gates for universal computation consists of~\cite{Weedbrook_RMP84_2012}: {\em{i)}} local Gaussian gates, such as the displacement gate $e^{-i \eta X}$, the phase gate $e^{-i \eta X^2}$, and the Fourier transform $e^{i \frac{\pi}{4}}e^{-i \frac{\pi}{4}\left(P^2+X^2\right)}$, {\em{ii)}} a non-local Gaussian gate that couples two different modes, like the CPHASE gate $e^{-i \eta X_i \otimes X_j}$, and {\em{iii)}} one local non-Gaussian gate, such as the cubic phase gate $e^{-i \eta X^3}$.
This example of the universal set of gates  is not unique; equivalent alternatives
can be considered. In optical systems, Gaussian gates can be relatively easily implemented using displacement, squeezing, phase shift, and beam splitter operations.
However, the implementation of non-Gaussian gates is much more difficult~\cite{Walschaers_PRX_2021}.

\section{Discrete representation of qumodes}
\label{sec:cvdvrep}

The representation of bosonic states on qubit hardware has been discussed in previous works~\cite{macridin_prl_2018,macridin_pra_2018,macridin_pra_2021,Li_PRA_2023}, with a focus on fermion-boson and scalar field quantum simulations. In this work, we briefly present the main ideas, emphasizing the points that are most relevant for  CV computations, and
qumode transfer protocols.

In this paper, we do not consider the direct encoding of Fock states to computational DV states. While this encoding efficiently represents states, we are not aware of any resource-efficient way to implement typical CV gates on DV devices using this encoding~\cite{macridin_pra_2021}.

\subsection{Nyquist-Shannon expansion of qumodes}
\label{sec:NSexp}

Let's start by assuming  that a cutoff $N_b$ can be chosen such that the contribution of states with more than $N_b$
bosons with mass $\mu$ is negligible to the qumode state, {\em i.e. $\ket{\phi} \approx \sum_{n=0}^{N_b} c_n \ket{n}$}, where $\ket{n}$ is the $n$ boson Fock state.
As can be seen from~\cref{eq:aop} the definition of boson operators is not unique;
the bosons are defined up to a mass factor, $\mu$.  Bosons with different masses
are related by a squeezing operation.
For a given qumode the cutoff $N_b$ depends on the boson mass $\mu$.
The smaller $N_b$ is, the better is the accuracy of the discrete representation of qumodes
which will be introduced in~\cref{sec:fhsp}. Keeping the boson mass as a tunable parameter can be useful for optimizing quantum algorithms and computational resources, as discussed in~\cite{macridin_pra_2021}. However, for the purpose of this paper, the boson mass $\mu$ is a fixed parameter.

The qumode's wavefunction $\phi(x)$  decreases exponentially fast to zero as the magnitude of its argument, $\abs{x}$, increases, since the Fock states' wavefunctions (Hermite-Gaussian functions) decrease exponentially fast to zero with increasing $\abs{x}$. The same is true for the wavefunction's Fourier transform; namely $\hphi(p)$ decreases exponentially fast to zero as the magnitude of its argument, $\abs{p}$, increases.
Therefore, for a desired accuracy $\epsilon$,
we can define a parameter $L_{\epsilon} >0$, as the \emph{minimum} value such that the weight of  $\phi(x)$ outside the interval $\left[-\frac{L_\e}{\sqrt{\mu}}, \frac{L_\e}{\sqrt{\mu}}\right]$ and the weight of $\hphi(p)$ outside the interval $\left[-L_\e\sqrt{\mu}, L_\e\sqrt{\mu}\right]$ are smaller or equal to $\epsilon$, {\em{i.e.}},
\begin{align}
\label{eq:tailwx}
\left( \int_{-\infty}^{ -\frac{L_\e}{\sqrt{\mu}} } \abs{\phi(x)}^2 dx + \int_{ \frac{L_\e}{\sqrt{\mu}} }^{\infty}  \abs{\phi(x)}^2 dx\right)^{\frac{1}{2}} &\le \epsilon~~\text{     and,} \\
\label{eq:tailwp}
\left( \int_{-\infty}^{ -L_\e\sqrt{\mu}}  \abs{\hphi(p)}^2 dp + \int_{ L_\e\sqrt{\mu}}^{\infty}  \abs{\hphi(p)}^2 dp\right)^{\frac{1}{2}} &\le \epsilon.
\end{align}
We call the intervals $\left[-\frac{L_\e}{\sqrt{\mu}}, \frac{L_\e}{\sqrt{\mu}}\right]$ and $\left[-L_\e\sqrt{\mu}, L_\e\sqrt{\mu}\right]$ the $\epsilon$-\emph{support intervals} of the functions $\phi(x)$ and $\hphi(p)$, respectively, since the functions are $\epsilon$-negligible for arguments outside those intervals.
The error $\epsilon$ decreases exponentially with increasing the support window parameter $L_\e$,
as analytical and numerical investigations reveal~\cite{macridin_pra_2021}.

The Nyquist-Shannon sampling theorem~\cite{Shannon_1949} states that a function with limited support in the Fourier space can be written as an infinite sum, with the sum terms proportional to the function sampled on a grid. In our case, the wavefunction is {\em{almost}} limited (\emph{i.e.} limited up to an error $\epsilon$) in {\em{both}} $x$ and $p$ variables. As a consequence, as discussed below, the wavefunction can be written up to an error $\e$ as a \emph{finite} sum, with the terms proportional
to the function sampled on a finite interval.

First, the Fourier transformed wavefunction $\hphi(p)$ is negligible ($\epsilon$-small) outside the interval $\left[-L_\e\sqrt{\mu}, L_\e\sqrt{\mu}\right]$. According to Nyquist-Shannon sampling theorem, this implies that the wavefunction $\phi(x)$ can be approximated by a discrete sampling such that
\begin{align}
\label{eq:NSphix}
\phi(x) = \sum_{j=-\infty}^{\infty} \phi(y_j) u(x-y_j)+\O(\epsilon),
\end{align}
\noindent where
\begin{align}
\label{eq:ygrid}
y_j&=\left(j+\delta\right) \Delta_{x}, \\
\label{eq:delta_x}
\Delta_{x}&=\frac{\pi}{L_\e\sqrt{\mu}},\\
\label{eq:sincu}
u(x)&=\text{sinc}\left(\frac{x}{\Delta_{x}}\right)
\equiv \frac{\sin\left(\pi \frac{x}{\Delta_{x}}\right)}{\pi \frac{x}{\Delta_{x}} }.
\end{align}
In~\cref{eq:NSphix} the term $\O(\epsilon)$ denotes a small quantity
with magnitude of the order $\epsilon$,
and is a consequence of the small weight of $\hphi(p)$ outside the widow $\left[-L_\e\sqrt{\mu}, L_\e\sqrt{\mu}\right]$.
In~\cref{eq:ygrid} $\delta \in \mathbb{R}$ is an arbitrary number, signifying that the Nyquist-Shannon expansions remains valid if the sampling grid is shifted by an arbitrary amount, as explained in~\cref{app:NShlf1}.
In \cref{eq:delta_x}, the discretization interval $\Delta_{x}$ is inversely proportional to the to the $\epsilon$-support interval in the Fourier space.

Second, according to~\cref{eq:tailwx}, the wavefunction $\phi(x)$ is $\epsilon$-small when $\abs{x}>\frac{L_\e}{\sqrt{\mu}}$.
Thus, the summation terms in~\cref{eq:NSphix} corresponding to the function sampled outside the
interval $\left[-\frac{L_\e}{\sqrt{\mu}}, \frac{L_\e}{\sqrt{\mu}}\right]$ can be neglected with an $\O(\epsilon)$ error.
The summation in~\cref{eq:NSphix} can be truncated to a finite sum with $N_\e$ terms and written as,
\begin{align}
\label{eq:phix}
\phi(x)= \sum_{j=0}^{N_\e-1} \phi(x_j+\delta_x \Delta_{x}) u(x-x_j-\delta_x \Delta_{x})+\O(\epsilon),
\end{align}
\noindent with
\begin{align}
\label{eq:grid_x}
x_j &= \left(j-\frac{N_\e-1}{2} \right)\Delta_{x},
\end{align}
\noindent and arbitrary $-0.5 < \delta_x \le 0.5$. The number of sampling points $N_\e$ is chosen
as the minimum number for which  the sampling points $\{x_j+\delta_x \Delta_{x}\}_{j \in \{0,...,N_\e-1 \} }$ covers the entire  sampling interval $\left[-\frac{L_\e}{\sqrt{\mu}}, \frac{L_\e}{\sqrt{\mu}}\right]$,  which implies
\begin{align}
\label{eq:Nxcond2}
x_0+\delta_x \Delta_x -\Delta_x&=\left(-\frac{N_\e+1}{2}+\delta_x\right) \Delta_x <- \frac{L_\e}{\sqrt{\mu}}\\
\label{eq:Nxcond1}
x_{N_\e-1}+\delta_x \Delta_x +\Delta_x&=\left(\frac{N_\e+1}{2}+\delta_x\right) \Delta_x > \frac{L_\e}{\sqrt{\mu}}.
\end{align}
\noindent \Cref{eq:delta_x,eq:Nxcond1,eq:Nxcond2} yield the relation between the
sampling interval parameter $L_\e$ and $N_\e$
\begin{align}
\label{eq:LNeps}
L_\e=\sqrt{\frac{\pi N_\e}{2}},
\end{align}
\noindent which is valid for all choices of $\abs{\delta_x} \le 0.5$. Note that the error term in~\cref{eq:phix} is larger than the one in~\cref{eq:NSphix} since it involves additional truncation approximation, although in both cases we denoted it as $\O(\epsilon)$. However, it is still of the order $\epsilon$.

A similar reasoning can be used to expand the wavefunction's Fourier transform $\hphi(p)$ by sampling it on the $N_\e$ points $\{p_m+\delta_p \Delta_{p}\}_{m \in \{0,...,N_\e-1 \} }$ covering the interval $\left[-L_\e\sqrt{\mu}, L_\e\sqrt{\mu}\right]$,
\begin{align}
\label{eq:phip}
\hphi(p) = \sum_{m=0}^{N_\e-1} \hphi(p_m+\delta_p \Delta_p) v(p-p_m-\delta_p \Delta_p)+\O(\epsilon)
\end{align}
\noindent where
\begin{align}
\label{eq:delta_p}
\Delta_{p}&=\frac{\pi\sqrt{\mu}}{L_\e}=\mu \Delta_x,\\
\label{eq:sincv}
v(p)&=\text{sinc}\left(\frac{p}{\Delta_{p}}\right),\\
\label{eq:grid_p}
p_m &= \left(m-\frac{N_x-1}{2} \right) \Delta_{p},
\end{align}
\noindent and $-0.5<\delta_p \le 0.5$ is an arbitrary shift.

The sampling sets $\{\phi(x_j+\delta_x \Delta_{x})\}_{j \in \{0,...,N_\e-1\}}$ and $\{\hphi(p_m+\delta_p \Delta_p)\}_{m\in \{0,...,N_\e-1\}}$ are connected by  shifted finite Fourier transforms, as follows:
\begin{align}
\label{eq:sh_dftphi}
\sqrt{\Delta_p} \hphi(p_m+\delta_p \Delta_p)&=\frac{1}{\sqrt{N_\e}}\sum_{j=0}^{N_\e-1} \sqrt{\Delta_x}\phi(x_j+\delta_x \Delta_{x}) e^{ -i \frac{2 \pi}{N_\e}\left(m-\frac{N_\e-1}{2}+\delta_p\right)  \left(j-\frac{N_\e-1}{2}+\delta_x\right)}+\O(\epsilon), \\
\label{eq:sh_dfthpi}
\sqrt{\Delta_x} \phi(x_j+\delta_x \Delta_{x})&=\frac{1}{\sqrt{N_\e}}\sum_{m=0}^{N_\e-1} \sqrt{\Delta_p}\hphi(p_m+\delta_p \Delta_p) e^{ i \frac{2 \pi}{N_\e}\left(m-\frac{N_\e-1}{2}+\delta_p\right)  \left(j-\frac{N_\e-1}{2}+\delta_x\right)}+\O(\epsilon).
\end{align}
\noindent \Cref{eq:sh_dftphi,eq:sh_dfthpi} can be derived by directly calculating the Fourier transforms of~\cref{eq:phix} and the inverse Fourier transform of~\cref{eq:phip}, respectively. Note that the Fourier
transform of the {\em{sinc}} function is the rectangular function, see~\cref{eq:ftu} in~\cref{app:sinc}.

\subsection{Finite Hilbert space representation}
\label{sec:fhsp}

For a given cutoff $N_b$, we construct a finite Hilbert space of dimension $N_x$, where $N_x$ is the number of sampling points necessary to discretize $\phi_{N_b}(x)$, the Fock wavefunction of order $N_b$,  with a desired accuracy $\e$. Thus, $N_x$ is given by
\begin{align}
\label{eq:LNx}
L=\sqrt{\frac{\pi N_x}{2}},
\end{align}
where $L\equiv L_\e (N_b)$ is the sampling interval parameter for $\phi_{N_b}(x)$
(see~\cref{eq:LNeps}).  For any $\epsilon<1$ the number of discretization points $N_x>N_b$,
as can be found by inspecting the properties of Fock state wavefunctions.

The finite Hilbert space is constructed by considering the basis $\{\ket{j}\}$ with $j \in \{0,1,\ldots, N_x-1\}$ and defining the discrete position operator $\X$ as,
\begin{align}
\label{eq:Xd}
\X\ket{j}=x_j\ket{j},
\end{align}
where $x_j$ is given by~\cref{eq:grid_x} with $N_\e$ and $L_\e$ replaced by $N_x$ and $L$, respectively. We also define the discrete momentum operator $\P$ as,
\begin{align}
\label{eq:Pd}
\P=\mu \tF \X \tF^{-1},
\end{align}
where $\tF$ represents the centered discrete Fourier Transform, defined by~\cref{eq:dft} in~\cref{app:DFT_sh} (see also~\cref{eq:dft1}).
The vectors $\{\ket{m}_p\}$, with $m \in \{0,1,\ldots, N_x-1\}$,
\begin{align}
\label{eq:dft1}
\ket{m}_p\equiv \tF\ket{m}= \frac{1}{\sqrt{N_x}}\sum_{j=0}^{N_x-1}    e^{i \frac{2 \pi}{N_x} \left(m-\frac{N_x-1}{2}\right) \left(j-\frac{N_x-1}{2}\right)}\ket{j}
\end{align}
\noindent are eigenvectors of $\P$,
\begin{align}
\label{eq:Pd1}
\P\ket{m}_p=p_m\ket{m}_p, ~\text{ with }~ p_m=\left(m-\frac{N_x-1}{2} \right)\Delta_p,
\end{align}
\noindent where
\begin{align}
\label{eq:Deltaxp}
\Delta_p=\frac{\pi\sqrt{\mu}}{L}=\mu \Delta_x.
\end{align}

The $\epsilon$-support intervals for the Fock wavefunction of order $N_b$ include the $\epsilon$-support intervals of all smaller order Fock wavefunctions, since the support interval parameter $L_\epsilon(n)$ monotonically increases with the Fock state order $n$~\cite{macridin_pra_2018}). This implies that, for $N_x$ discretization points, the discretization errors of all $n<N_b$ Fock states are smaller than $\O(\epsilon)$.
Then, for all $n<N_b$ the vectors defined by
\begin{align}
\label{eq:nvec}
\ket{\widetilde{n}}\equiv\sqrt{\Delta_x}\sum_{j=0}^{N_x-1} \phi_n(x_j)\ket{j}=
\sqrt{\Delta_p}\sum_{m=0}^{N_x-1} \hphi_n(p_m)\ket{m}_p+\O(\epsilon),
\end{align}
\noindent where $\phi_n(x)$ is the $n$-Fock state's wavefunction, satisfy
\begin{align}
\label{eq:npropx}
\X\ket{\widetilde{n}}&=\frac{1}{\sqrt{2 \mu}}\left(\sqrt{n}\ket{\widetilde{n-1}}+\sqrt{n+1}\ket{\widetilde{n+1}} \right)+\O(\epsilon)\\
\label{eq:npropp}
\P\ket{\tilde{n}}&=-i\sqrt{\frac{\mu}{2 }}\left(\sqrt{n}\ket{\widetilde{n-1}}-\sqrt{n+1}\ket{\widetilde{n+1}} \right)+\O(\epsilon).
\end{align}
\noindent  \Cref{eq:npropx,eq:npropp} can be obtained by employing~\cref{eq:Xd,eq:Pd1}
and the following properties of the Hermit-Gaussian functions: $x\phi_n(x)=\frac{1}{\sqrt{2 \mu}}\left[\sqrt{n}\phi_{n-1}(x)+\sqrt{n+1}\phi_{n+1}(x)\right]$ and
$p\hphi_n(p)=-i\sqrt{\frac{\mu}{2 }}\left[\sqrt{n}\hphi_{n-1}(p)-\sqrt{n+1}\hphi_{n+1}(p)\right]$, respectively.

Employing~\cref{eq:npropx,eq:npropp} it can be shown that the vectors
$\{\ket{\widetilde{n}}\}_{n<N_b}$ are, up to an error of order $\O(\epsilon)$,
the eigenvectors of the discrete harmonic oscillator, \emph{i.e.}
\begin{align}
\label{eq:dhoeigen}
H_h\ket{\widetilde{n}}=\mu\left(n+\frac{1}{2}\right)\ket{\widetilde{n}}+\O(\epsilon),
\end{align}
\noindent where
\begin{align}
\label{eq:dho}
H_h=\frac{1}{2}\P^2+ \frac{\mu^2}{2}\X^2.
\end{align}
\noindent Analogously it can be shown that
\begin{align}
\label{eq:dxpcom0}
\left(\X \P-\P \X\right)\ket{\widetilde{n}}=i\ket{\widetilde{n}}+\O(\epsilon),
\end{align}
\noindent for $n<N_b$.

In other words, on the $N_b$ dimensional subspace defined by the projector $\Q_b$
\begin{align}
\label{eq:qbdef}
\Q_b =\sum_{n=0}^{N_b-1} \ket{\n}\bra{\n},
\end{align}
\noindent where  $\{\ket{\n}\}_n$ are the eigenvectors of the discrete harmonic oscillator defined by~\cref{eq:dho},
the discrete position and momentum operators obey
(up to an error term $\O(\epsilon)$) the canonical commutation relation, \emph{i.e.}
\begin{align}
\label{eq:dxpcom}
\left[\X, \P\right]\Q_b=i\Q_b+\O(\epsilon).
\end{align}

Let
\begin{align}
\label{eq:qbCdef}
Q_b=\sum_{n=0}^{N_b-1} \ket{n}\bra{n},
\end{align}
\noindent with  $\ket{n}$ being the $n$-th Fock state of the CV Hilbert space, denoting the projector on the subspace with the number of bosons below $N_b$.
As can be seen from~\cref{eq:npropx,eq:npropp}, the operators $\X$ and $\P$ acts on the
subspace projected by $\Q_b$ as the operators $X$ and, respectively, $P$ acts on the subspace of the continuous  Hilbert space defined by the projector $Q_b$.
There is an isomorphism between the CV subspace defined by the projector $Q_b$ and the subspace of the finite Hilbert space defined by $\Q_b$ (for illustration see Fig.~$2$ in~\cite{macridin_pra_2021}). A CV wavefunction characterizing a qumode with less than $N_b$ bosons can be encoded with $\O(\epsilon)$ error on the discrete system of size $N_x$ as follows:
\begin{align}
\label{eq:cvdvrep}
\ket{\phi_C}=\int \phi(x) \ket{x}_C dx \longleftrightarrow \ket{\phi_D} = \sqrt{\Delta_x} \sum_{j=0}^{N_x-1} \phi(x_j) \ket{j}_D.
\end{align}
Furthermore, a CV operator $O(X,P)$ generated by $X$ and $P$ that acts on and yields states in the subspace defined by the projector $Q_b$ can be mapped to the operator $\bar{O}(\X,\P)$ which acts on the discrete space, by replacing $X$ and $P$ with $\X$ and $\P$, respectively:
\begin{align}
\label{eq:opmap}
O(X,P) Q_b \longleftrightarrow  \bar{O}(\X,\P) \Q_b~\text{ when }~O(X,P)Q_b = Q_b O(X,P)Q_b+\O(\epsilon).
\end{align}

By inspecting~\cref{eq:phix}, it is clear that the information encoded in the DV state $\ket{\phi_D}$, as described by~\cref{eq:cvdvrep}, is sufficient to reproduce (up to an error $\O(\e)$) the CV wavefunction $\phi(x)$ for all values of $x$.  In fact,  $\phi(x)$ at a particular $x$ can be directly measured in the DV basis $\{\ket{j}\}$ by applying the grid shift operator $T_{\delta,0}$ before the measurement,
\begin{align}
\label{eq:Tshift}
T_{\delta,0} \left[\sqrt{\Delta_x} \sum_{j=0}^{N_x-1} \phi(x_j) \ket{j}\right]=\sqrt{\Delta_x} \sum_{j=0}^{N_x-1}
\phi(x_j+\delta\Delta_x) \ket{j}+ \O(\epsilon),
\end{align}
where $\delta=\frac{(x-x_l)}{\Delta_x}$ and $x_l$ is the grid point closest to $x$. The grid shift operator $T_{\delta,0}$ is a product of a shifted Fourier transform with an inverse shifted Fourier transform and is defined in~\cref{eq:top} in~\cref{app:DFT_sh}.

\subsection{Finite Hilbert space encoding on qubits}
\label{sec:qmodeonqubits}

The $N_x$ basis states $\{\ket{j}\}$, with integer $j \in \{0,...,N_x-1\}$ are represented on $n_q=\log_2(N_x)$ qubits in a binary encoding
\begin{align}
\label{eq:xjqub}
\ket{j}=\ket{j_0}\ket{j_1}...\ket{j_{n_q-1}},
\end{align}
\noindent where $j_q \in \{0,1\}$, such that
\begin{align}
\label{eq:jqub}
j=\sum_{q=0}^{n_q-1}j_q2^{n_q-1-q}.
\end{align}

The discrete position operator is expressed  as
\begin{align}
\label{eq:Xdqub}
\X= -\Delta_x \sum_{q=0}^{n_q-1}2^{n_q-1-q} \frac{\sigma^z_q}{2},
\end{align}
\noindent where $\sigma^z_q=\ket{0}\bra{0}_q-\ket{1}\bra{1}_q$ is the Pauli
$\sigma^z$ acting on the qubit $q$. The operator $\X$ satisfies~\cref{eq:Xd},
as can be directly checked.

The implementation of the discrete momentum operator $\P$ is achieved by using~\cref{eq:Pd}, along with the implementation of the centered discrete Quantum Fourier transform described in~\cref{sec:sftqubits}.

The gates required for universal CV quantum computation can be implemented on qubits by replacing $X$ and $P$ with $\X$ and $\P$ respectively, as mentioned in~\cref{sec:fhsp}. In~\cref{sec:gates}, we present the explicit implementation on qubits of the universal set of gates introduced in~\cref{sec:cv}. The number of elementary single-qubit and two-qubit gates required for this implementation scales polynomially with the size of the DV device. This is one of the main advantages of our encoding scheme: the CV gates can be resource-efficiently implemented on qubit hardware.

Additionally, in~\cref{sec:squeezing}, we  provide an implementation of the discrete squeezing operator,
\begin{align}
\label{eq:dsqeeze}
\S(r)=e^{ i \frac{r}{2}\left(\X\P+\P\X\right)}.
\end{align}
\noindent The discrete squeezing operator  will be used in~\cref{sec:discard_qubits,sec:add_qubits} to discard or add ancillary qubits to the DV device in order to increase the transfer protocol success probability.
For that we will use  the following property of $\S(r)$,
\begin{align}
\label{eq:squeezephi}
\S(r)\sqrt{\Delta_x} \sum_{j=0}^{N_x-1} \phi(x_j) \ket{j}=\sqrt{\Delta_x e^{{r}}} \sum_{j=0}^{N_x-1} \phi(x_j e^{r}) \ket{j}+\O(\epsilon),
\end{align}
\noindent valid when both the initial and the squeezed qumode have
negligible weight on the subspace with more than $N_b$ bosons.

\subsection{Qumode representation errors}
\label{sec:errors}

Detailed analytical and numerical investigations of the errors encountered during the construction of the finite representation of the continuous Hilbert space are presented in~\cite{macridin_pra_2021}.
For a fixed error $\epsilon$, we find numerically that the number of discretization points $N_x$ is approximately proportional to the cutoff $N_b$ ($N_x \approx c_1 + c_2N_b$ where $c_1$ and $c_2$ are dependent on $\epsilon$).
We find that the ratio $N_b/N_x$ falls within the range of $\left[0.3, 0.7\right]$ when the error is in the range of $\left[10^{-5}, 10^{-3}\right]$.
For a given cutoff $N_b$, the error $\epsilon$ decreases exponentially as the number of discretization points increases, because the support interval parameter $L \propto \sqrt{N_x}$ and $\e$ decreases exponentially with increasing $L$. To give some examples, we find numerically that DV devices with $n_q=6$ and $7$ qubits can represent qumodes with cutoffs of $N_b=30$ and $70$, respectively, with an accuracy of $\epsilon=10^{-4}$.

The construction of the finite representation for a cutoff $N_b$ and the error analysis discussed so far, assume that the cutoff $N_b$ can be chosen such that the contribution of Fock states with more than $N_b$ bosons is negligible.
However the errors introduced by the truncation in the Fock basis also need to be considered. This error  can be quantified by
\begin{align}
\label{eq:wnb}
\omega_{N_b}=\norm{\left(\mathbb{1}-Q_b\right)\ket{\phi}}=\sqrt{\sum_{n=N_b}^{\infty} \abs{\braket{n}{\phi}}^2},
\end{align}
\noindent where $Q_b$ is the projector on the subspace with less than $N_b$ bosons (see~\cref{eq:qbCdef}).


For a desired error of order $\epsilon$ in the qumode discrete representation, the cutoff $N_b$ should be chosen so that $\omega_{N_b} \approx \epsilon$, and the number of discretization points $N_x$ should be chosen so that the Fock state of order $N_b$ is discretized with an error of order $\epsilon$.

A relevant question for managing computational resources is how the number of required qubits scales with the error $\epsilon$. Depending on the behaviour of the qumode boson distribution at large
$n$, there are two cases to be discussed:

\emph{i)} The qumode boson truncation error $\omega_{N_b}$ decreases faster than exponentially with increasing $N_b$. In this case, for a choice of $N_b$ large enough, $\omega_{N_b}$ becomes negligible, and the error will be dominated by the quadrature (\emph{i.e.} position and momentum) discretization error of the Fock states with orders smaller than $N_b$. As previously discussed, these errors decrease exponentially with an increase in the number of discretization points, \emph{i.e.}, the required number of qubits scales as $n_q = \log_2 N_x \propto \log_2 \left[ \log \left(\epsilon^{-1}\right) \right]$.

Note that this case includes coherent states, displaced number states, and squeezed states, albeit for large displacements or strong squeezing the cutoff $N_{b}$ is large. For these states, the probability of having $n$ bosons is bounded by  $P_n \propto \frac{C^n}{n!}$~\cite{gerry_knight_2004}, where $C$ is independent on $n$ and determined by the displacement parameter or, respectively, the squeeze parameter. By employing the Stirling formula $n! \approx \sqrt{2\pi n} \left(e/n\right)^n$
one can see that for $n \gg eC$ the boson distribution and $\omega_{N_b}$ decrease faster than exponentially with increasing $n$.

\emph{ii)} The qumode boson truncation error $\omega_{N_b}$ decreases slower
then exponentially with increasing $N_b$. In this case, the dominant
error will be $\epsilon(N_b) \approx \omega_{N_b}$. The number of qubits scales as $n_q = \log_2 N_x \propto \log_2 \left[N_b(\epsilon)\right]$, where $N_b(\epsilon)$ is obtained by solving the equation $\epsilon = \omega_{N_b}$. This case includes the situations where the
qumode's wavefunction in the $\{\ket{x}\}$ basis, or $\{\ket{p}\}$ basis,  decreases slower than exponentially with increasing $\abs{x}$, or $\abs{p}$, respectively.

In practice, for both cases, the accuracy of the approximation can be controlled by increasing the number of qubits until the results are converged within the desired error.


\section{Transfer protocols}
\label{sec:teleprotocols}

In this section we introduce two transfer protocols. Both protocols are modification of the {\em{one-qubit}} CV teleportation protocol described in~\cite{Menicucci_PRL97_2006,Weedbrook_RMP84_2012}.

The goal of the first  protocol is to  transfer a CV qumode
\begin{align}
\label{eq:cvstate}
\ket{\phi_C}=\int  \phi(x) \ket{x}_C dx=\int  \sum_{j=0}^{N_x-1} \phi(x_j) u(x-x_j) \ket{x}_C dx+err(N_x),
\end{align}
\noindent to its discrete representation
\begin{align}
\label{eq:dvstate}
\ket{\phi_D}= \sqrt{\Delta_x} \sum_{j=0}^{N_x-1} \phi(x_j) \ket{j}_D.
\end{align}
\noindent Note that in~\cref{eq:cvstate}, unlike in the previous sections, we denote the error term arising from the qumode discretization as $err(N_x)$. In the following  we will use $\e$ to denote the error of the transfer protocols.

We measure the protocol fidelity by
\begin{align}
\label{eq:teleFCD}
F_D=\abs{ \braket{\chi_D}{\phi_D} },~\text{ where }~\ket{\chi_D}=\T^{CD}\left(\ket{\phi_C}\right),
\end{align}
\noindent and $\T^{CD}$ represents the transfer channel taking a CV state to a DV device.

The goal of the second transport protocol is to take the DV state described by ~\cref{eq:dvstate} to the corresponding CV state described by~\cref{eq:cvstate}. The  fidelity for this protocol
is \begin{align}
\label{eq:teleFDC}
F_C=\abs{ \braket{\chi_C}{\phi_C} },~\text{ where }~\ket{\chi_C}=\T^{DC}\left(\ket{\phi_D}\right),
\end{align}
\noindent and $\T^{DC}$ represents the transfer channel taking a DV state to a CV device.

The transfer protocols involve measurement operations and the resulting fidelity is dependent on the measurement outcome. We consider the transfer successful if the  fidelity is larger than a desired threshold value. As described in sections \cref{sec:tele_cvdv,sec:tele_dvcv}, the success of the protocols is conditioned on the measurement outcome, and for certain outcomes, the transfer fails. To quantify the success of the protocols, we define the \emph{transfer probability of success} as the probability that the measurement outcome belongs to the set of measurements that yield a successful transfer. As can be inferred from the above definition, the transfer probability of success is dependent on the chosen fidelity threshold.

In the following, we will denote by $L_{\epsilon}$ the parameter that determines the $\epsilon$-support intervals for the qumode's wavefunction (defined by~\cref{eq:tailwx,eq:tailwp}). We will also have the parameter $L=\sqrt{\pi N_x/2}$ (as in~\cref{eq:LNx}), determined by the number of qumode discretization points $N_x$. As discussed in the next sections, for both transfer protocols, in order to achieve a significant probability of success with a desired error $\e$,
$N_x$ needs to be chosen large enough so that $L \gg L_{\epsilon}$. In this case, the term $err(N_x)$ in ~\cref{eq:cvstate} will be much smaller than $\epsilon$ (since $err$ decreases fast with increasing $L$) and, since it is not the dominant error, it will be neglected
in the following analysis of the transfer protocol.

\subsection{Coupling between continuous-variable and discrete-variable devices}
\label{sec:couplingCD}

To implement the transfer protocols, we assume that the unitary
\begin{align}
\label{eq:cpaseCD}
e^{-i \eta X \otimes \X},
\end{align}
coupling the CV and DV devices can be implemented. Since $\X$ is a linear combination of $\sigma^z_q$ operators (see \cref{eq:Xdqub}), this can be achieved if the unitary $e^{-i \eta X \otimes \sigma^z_q}$ coupling the qumode and the qubit $q$ can be realized for all $q \in \{0,1,...,n_q-1\}$. For example,  this type of mode-qubit coupling can be achieved by considering the evolution under the interaction Hamiltonian $H_{int} \propto \left(a^\dagger + a\right) \sigma^x_q$ sandwiched between two qubit Hadamard gates
\begin{align}
e^{-i \eta X \otimes \sigma^z_q} =H_q e^{-i \eta X \otimes \sigma^x_q} H_q.
\end{align}
\noindent This kind of interaction is realized, for instance, in systems with transmons coupled to a microwave cavity \cite{Bishop2009}, or in systems with an electromagnetic mode coupled to qubits \cite{Walther2006, Cottet2017, Blais2021}.

\subsection{Qumode transfer from CV device to DV device}
\label{sec:tele_cvdv}

\begin{figure}[tb]
    \begin{center}
        \includegraphics*[width=5in]{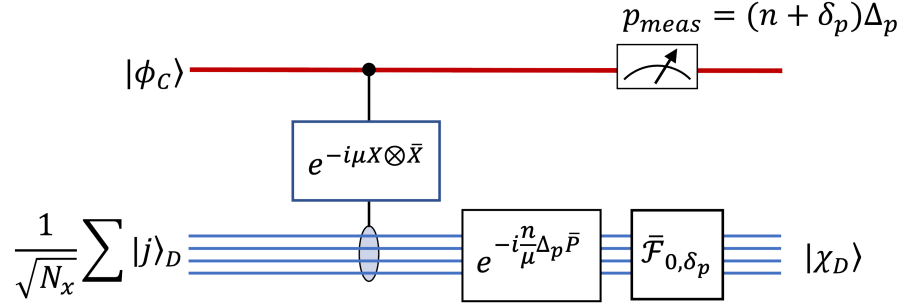}
        \caption{CV-DV transfer protocol, described in~\cref{sec:tele_cvdv}. The CV state
        $\ket{\phi_C}$ is transferred into the DV state $\ket{\chi_D}$.}
        \label{fig:diagcvdv}
    \end{center}
\end{figure}

The CV-DV transfer protocol, diagrammatically presented in~\cref{fig:diagcvdv},  consists of the following steps:

\begin{enumerate}
\item
By applying a  Hadamard gate to every qubit, the  DV system is prepared into the state
\begin{align}
\label{eq:dvini}
\frac{1}{\sqrt{N_x}}\sum_{j=0}^{N_x-1}  \ket{j}_D.
\end{align}
\noindent The initial joint CV-DV system's state is
\begin{align}
\label{eq:cdini}
\ket{\chi_{CD}}=\frac{1}{\sqrt{N_x}}\int  \sum_{j=0}^{N_x-1} \phi(x) \ket{x}_C \ket{j}_D dx.
\end{align}

\item The entangling operator $e^{-i \mu X \otimes \X }$ is applied.  The state becomes
\begin{align}
\label{eq:tele_cvdv_1}
e^{-i \mu X \otimes \X }\ket{\chi_{CD}}&=\frac{1}{\sqrt{N_x}}\int  \sum_{j=0}^{N_x-1} e^{-i \mu x x_j }\phi(x) \ket{x}_C \ket{j}_D dx.
\end{align}

\item The CV system is measured in the momentum basis by employing a  homodyne measurement.
Let's denote the measurement result  by $p_{meas}$. After the measurement, the  DV state  becomes
\begin{align}
\label{eq:pstateD}
\ket{\chi_{D0}}&=\frac{1}{\sqrt{\text{Pr}(p_{meas})}}\frac{1}{\sqrt{2 \pi N_x}}\int  \sum_{j=0}^{N_x-1} e^{-i x \left(\mu x_j+p_{meas}\right) }\phi(x)  \ket{j}_D dx\\ \nonumber
&= \frac{1}{\sqrt{\text{Pr}(p_{meas})}}\frac{1}{\sqrt{N_x}}\sum_{j=0}^{N_x-1} \hphi(\mu x_j+p_{meas}) \ket{j}_D,
\end{align}
\noindent while the probability to measure the value $p_{meas}$ is
\begin{align}
\label{eq:propb}
\text{Pr}(p_{meas})&= \frac{1}{N_x} \sum_{j=0}^{N_x-1} \abs{\hphi(\mu x_j+p_{meas})}^2.
\end{align}

\item The gate $e^{-i \frac{n \Delta_p }{ \mu}\P }$ is applied to the DV device,
\begin{align}
\label{eq:pstate2}
\ket{\chi_{D1}}&=e^{-i \frac{n \Delta_p}{ \mu}\P }\ket{\chi_{D0}}= \frac{1}{\sqrt{\text{Pr}(p_{meas})}}\frac{1}{\sqrt{N_x}}\sum_{j=0}^{N_x-1} \hphi \left[\mu x_j+\left(n+\delta_p\right) \Delta_p\right] \ket{ \left(j+n\right)_{mod N_x}}_D,
\end{align}
\noindent where the integer $n$ and the shift parameter $-0.5<\delta_p\le 0.5$ are defined such that $p_{meas}=\left(n +\delta_p\right) \Delta_p$.
Here $k_{mod N_x}=:k-N_x \lfloor\frac{k}{N_x}\rfloor$, with $\lfloor~\rfloor$ being the integer {\em{floor}} function,  denotes $k \textit{ modulo } N_x$  and takes integer values  between $0$ and $N_x-1$.

\item For the final step the shifted Fourier transform (see~\cref{eq:sft} in~\cref{app:DFT_sh}),
\begin{align}
\tF_{0,\delta_p}=\frac{1}{\sqrt{N_x}} \sum_{l,j=0}^{N_x-1}    e^{i \frac{2 \pi}{N_x}
\left(l - \frac{N_x-1}{2}\right) \left(j- \frac{N_x-1}{2}+\delta_p\right)}  \ket{l}_D\bra{j}_D,
\end{align}
\noindent is applied to the DV state. The transferred state is
\begin{align}
\label{eq:TCDchi}
\T^{CD}(\ket{\phi_C})\equiv\ket{\chi_D}=\tF_{0,\delta_p}\ket{\chi_{D1}}=  \sum_{j=0}^{N_x-1} \xi_j\ket{j}_D,
\end{align}
\noindent where
\begin{align}
\label{eq:TCDphij}
\xi_j= \frac{1}{N_x\sqrt{\text{Pr}(p_{meas})}}   \sum_{l=0}^{N_x-1}  \hphi \left(\mu x_{l+n}+\delta_p\Delta_p\right)
e^{i \frac{2 \pi}{N_x} \left(j - \frac{N_x-1}{2}\right)\left[\left(l + n\right)_{mod N_x}-\frac{N_x-1}{2}+\delta_p\right]}.
\end{align}

\end{enumerate}

\begin{figure}[tb]
    \begin{center}
        \includegraphics*[width=5in]{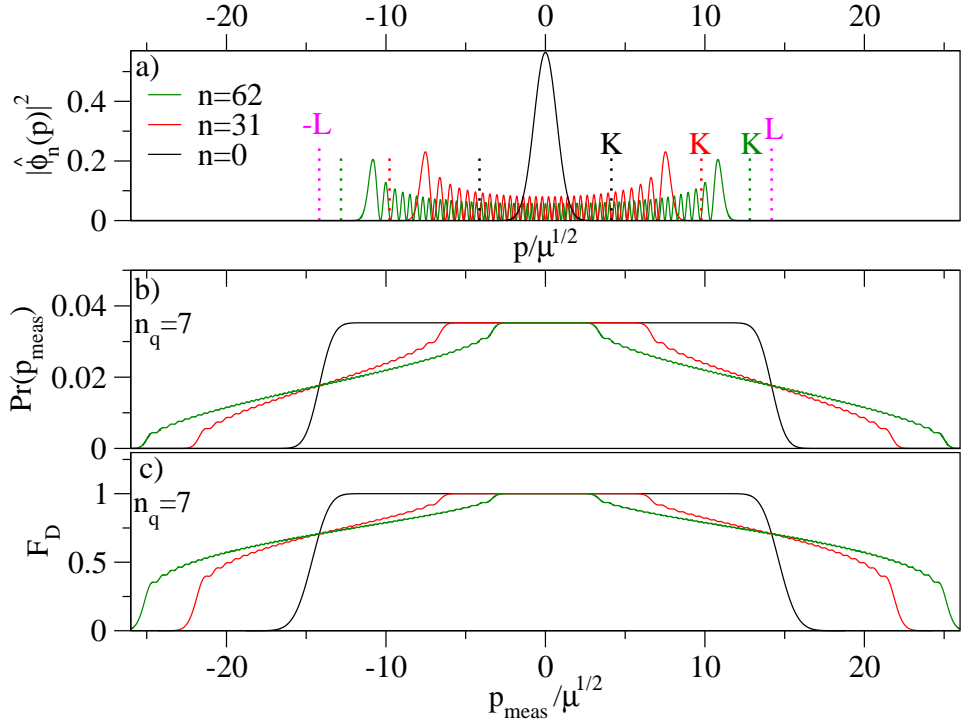}
        \caption{CV-DV transfer protocol of $n=0$, $n=31$ and $n=62$ Fock states (Hermite-Gaussian functions)
        to a $n_q=7$ qubit device. a) The momentum distribution $\abs{\hphi_n(p)}^2$ of the Fock states. Here $L=\sqrt{\frac{\pi 2^{n_q}}{2}}$ and  $L_{\epsilon}(n)$ is defined by the~\cref{eq:tailwx,eq:tailwp} for $\epsilon=10^{-4}$. $L_{\epsilon}$ increases with increasing the order $n$ of the Fock state.
        b) The probability to measure the $p_{meas}$.
        c) The transfer fidelity $F_D$  versus $p_{meas}$. The  fidelity
        $F_D \ge 1-\O(\epsilon)$  when $p_{meas} \in [-L\sqrt{\mu}+L_{\epsilon}\sqrt{\mu}-\frac{\Delta_p}{2}, L\sqrt{\mu}-L_{\epsilon}\sqrt{\mu}+\frac{\Delta_p}{2}]$.
        }
        \label{fig:telecvdv}
    \end{center}
\end{figure}

As can be seen from~\cref{eq:TCDphij} the transfer protocol fidelity depends on  the value of
$p_{meas}=\left(n +\delta_p\right) \Delta_p$. We will show that for the values of $p_{meas}$ for which
\begin{equation}
\label{eq:tcond}
\hphi \left(\mu x_{j+n}+\delta_p \Delta_p\right) = \hphi \left(\mu x_{(j+n)_{mod N_x}}+\delta_p \Delta_p\right)+\O(\epsilon)~\text{ for all }~j \in \{0,...,N_x-1\},
\end{equation}
\noindent  the transfer protocol has a small error $\O(\epsilon)$.

First, we determine  $p_{meas}$ for which~\cref{eq:tcond} is true.
There are three cases to be discussed:

{\em{i)}} When $j+n=(j+n)_{mod N_x}$, ~\cref{eq:tcond} is obviously satisfied.

{\em{ii)}} For positive $n$, when
$j+n \ge N_x$, the modulo sum  $(j+n)_{mod N_x}=j+n-N_x$. On the left-hand side of ~\cref{eq:tcond}
we have $\hphi \left(\mu x_{j+n}+\delta_p \Delta_p\right)=\O(\epsilon)$ since  $\mu x_{j+n}+\delta_p \Delta_p>L_{\epsilon}\sqrt{\mu}$ is outside the $\epsilon$-support interval of the $\hphi$ function.
The requirement that the right-hand side
of~\cref{eq:tcond}  $\hphi \left(\mu x_{(j+n)_{mod N_x}}+\delta_p \Delta_p\right)=\O(\epsilon)$ implies $\mu x_{j+n-N_x}+\delta_p \Delta_p< -L_{\epsilon}\sqrt{\mu}$. This is equivalent to $\left(j+n-N_x-\frac{N_x-1}{2}+\delta_p\right) \Delta_p< -L_{\epsilon}\sqrt{\mu}$ for $j =N_x-1$ and, by employing~\cref{eq:LNx}, implies $(n+\delta_p)\Delta_p<-L_{\epsilon} \sqrt{\mu}
+L\sqrt{\mu}+\frac{\Delta_p}{2}$.

{\em{iii)}} For negative $n$, when
$j+n<0$, the modulo sum  $(j+n)_{mod N_x}=j+n+N_x$. The left-hand side of~\cref{eq:tcond} $\hphi \left(\mu x_{j+n}+\delta_p \Delta_p\right)=\O(\epsilon)$ since the argument $\mu x_{j+n}+\delta_p \Delta_p<-L_{\epsilon}\sqrt{\mu}$ is outside the $\epsilon$-support interval of the $\hphi$ function. The requirement that the right-hand side of~\cref{eq:tcond}  $\hphi \left(\mu x_{(j+n)_{mod N_x}}+\delta_p \Delta_p\right)=\O(\epsilon)$ implies $\mu x_{j+n+N_x}+\delta_p \Delta_p> L_{\epsilon}\sqrt{\mu}$. This is equivalent to $\left(j+n+N_x-\frac{N_x-1}{2}+\delta_p\right) \Delta_p> L_{\epsilon}\sqrt{\mu}$ for $j=0$ which implies $(n+\delta_p)\Delta_p>L_{\epsilon} \sqrt{\mu}
-L\sqrt{\mu}-\frac{\Delta_p}{2}$.

Considering {\em{i)}}, {\em{ii)}} and {\em{iii)}}, we can conclude that~\cref{eq:tcond} is satisfied when
\begin{align}
\label{eq:pvalid}
\abs{p_{meas}} <L\sqrt{\mu}-L_{\epsilon}\sqrt{\mu}+\frac{\Delta_p}{2}.
\end{align}

Second, we calculate the probability to measure $p_{meas}$
when $p_{meas}$ satisfies \cref{eq:pvalid}.
In~\cref{eq:propb} the argument  $\mu x_j+p_{meas}$ of the wavefunction $\hphi(p)$
takes values in the interval $[ -L\sqrt{\mu}-\frac{1}{2}\Delta_p+p_{meas},L\sqrt{\mu}+\frac{1}{2}\Delta_p+p_{meas}]$ when the summation index $j$ runs from $0$ to $N_x-1$.
The summation terms for which $\mu x_j+p_{meas}$ takes value outside the $\epsilon$-support interval $\left[ -L_{\epsilon}\sqrt{\mu}, L_{\epsilon}\sqrt{\mu}\right]$ are negligible small (of order $\O(\epsilon)$) and, therefore, their contribution to the sum is negligible. In another words,  as long as $\left[ -L_{\epsilon}\sqrt{\mu},  L_{\epsilon}\sqrt{\mu}\right] \subset [ -L\sqrt{\mu}-\frac{1}{2}\Delta_p+p_{meas},L\sqrt{\mu}+\frac{1}{2}\Delta_p+p_{meas}]$ (which is equivalent to \cref{eq:pvalid}),
all the non-negligible terms are included in the summation. In this case the sum is independent
of the value of $p_{meas}$, {\em{i.e.}},
\begin{align}
\label{eq:probmeas}
\text{Pr}(p_{meas}) = \frac{1}{N_x} \sum_{j=0}^{N_x-1} \abs{\hphi(\mu x_j)}^2 +\O(\epsilon)
=\frac{1}{N_x \Delta_p}+\O(\epsilon)
~\text{ for }~ \abs{p_{meas}} <L\sqrt{\mu}-L_{\epsilon}\sqrt{\mu}+\frac{\Delta_p}{2}.
\end{align}
\noindent The last equality in~\cref{eq:probmeas} results from the normalization of $\hphi(p)$ when~\cref{eq:phip} and the orthogonality properties of the \emph{sinc} functions (\cref{eq:ortho})  are employed.

Finally, for $p_{meas}$ satisfying~\cref{eq:pvalid}, ~\cref{eq:sh_dftphi,eq:tcond,eq:probmeas,eq:TCDphij}
imply
\begin{align}
\label{eq:xij}
\xi_j= \sqrt{\Delta_x}\phi(x_j) +\O(\epsilon).
\end{align}
\noindent  \Cref{eq:TCDchi,eq:xij} imply that
 \begin{align}
\label{eq:cvdvtele}
\T^{CD}(\ket{\phi_C})=\ket{\phi_D}+\O(\epsilon) ~\text{ for }~ \abs{p_{meas}} <L\sqrt{\mu}-L_{\epsilon}\sqrt{\mu}+\frac{\Delta_p}{2}.
\end{align}
\noindent Hence, for $p_{meas}$ in the interval range given by~\cref{eq:pvalid}, the CV-DV transfer operation has a small error $\O(\epsilon)$.

\begin{figure}[tb]
    \begin{center}
        \includegraphics*[width=5in]{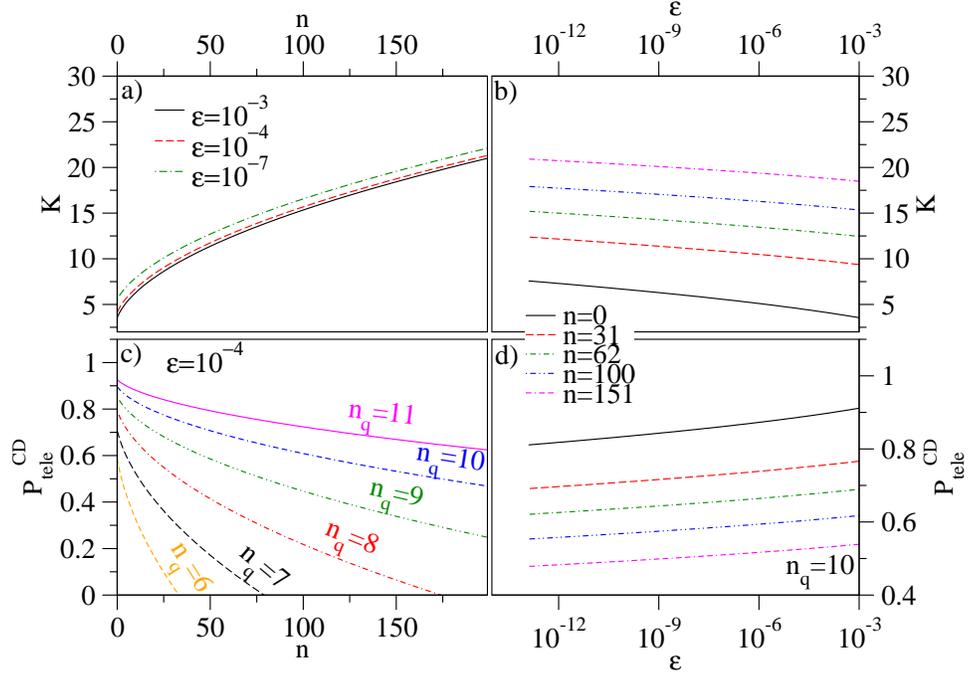}
        \caption{a) The support interval parameter $L_{\epsilon}$ (dimensionless) defined by~\cref{eq:tailwx,eq:tailwp}
        versus the Fock state order $n$ for different values of the error $\epsilon$. Numerical fitting
        yields $L_{\epsilon} \propto \sqrt{2n+c_1}+c_2$, where $c_1$ and $c_2$ are constants of the order of unity and depend on $\epsilon$.
        b)  $L_{\epsilon}$ versus the error $\epsilon$ (logarithmic scale) for Fock state
        with $n=0, 31, 62, 100$ and $151$. Numerical fitting finds that the error decreases exponentially with increasing $L_{\epsilon}$, \emph{i.e.}, $\epsilon \propto e^{-c L_{\epsilon}}$ where $c \approx 10$.
        c) Probability $P^{CD}$ (see \cref{eq:probCD_eps}) for high fidelity transfer ($\epsilon=10^{-4}$) versus $n$ for
        DV devices with different number of qubits.  $P^{CD}$ decreases with increasing $n$ since $L_{\epsilon}$ increases with increasing $n$. $P^{CD}$ increases with increasing $n_q$ since $L$ increases with increasing $n_q$.
        d) Probability $P^{CD}$ versus $\epsilon$ (logarithmic scale) for Fock state
        with $n=0, 31, 62, 100$ and $151$ for a DV device with $n_q=10$ qubits.
        }
        \label{fig:pgood}
    \end{center}
\end{figure}

The protocol probability of success, defined as the probability of having a measurement outcome such that the fidelity is larger than $1-\epsilon$, is given by
\begin{align}
\label{eq:probCD_eps}
P^{CD}(\epsilon)= \int dp_{meas} \text{Pr}(p_{meas})\bigg|_{F_D>1-\epsilon},
\end{align}
\noindent where the fidelity $F_D$ is defined by~\cref{eq:teleFCD}.
 According to ~\cref{eq:probmeas,eq:cvdvtele} we have
\begin{align}
\label{eq:probCD_tele}
P^{CD}(\epsilon) \approx P^{CD}\left[\O(\epsilon)\right]
= \int_{-\left(L-L_{\epsilon}\right) \sqrt{\mu}-\frac{\Delta_p}{2}}^{\left(L-L_{\epsilon}\right) \sqrt{\mu}+\frac{\Delta_p}{2}} \text{Pr}(p_{meas}) dp_{meas}
=\frac{L-L_{\epsilon}}{L}+\frac{1}{N_x}=\frac{\sqrt{N_x}-L_{\epsilon}\sqrt{2/\pi}}{\sqrt{N_x}}+\frac{1}{N_x}.
\end{align}
Note that a high probability of success requires $L \gg L_\e$, which implies that the wavefunction discretization error $err(N_x) \ll \e$ (see~\cref{eq:cvstate}). This justifies the omission
of terms of order $err(N_x)$ in our description of the transfer protocols.

\Cref{eq:probCD_tele} shows that  the probability of a successful transfer protocol increases with increasing the number of discretization points,
\begin{align}
\label{eq:prob_good_assim}
P^{CD}(\epsilon) \xrightarrow{N_x \rightarrow \infty} 1.
\end{align}
\noindent  Considering that the number of the discretization points increases exponentially with
number of qubits ($N_x=2^{n_q}$),~\cref{eq:probCD_tele} implies
that the probability of failure decreases exponentially with increasing the number of qubits,
\begin{align}
\label{eq:prob_failure}
1-P^{CD}(\epsilon)=L_{\epsilon}\sqrt{\frac{2}{\pi}} 2^{-\frac{n_q}{2}}-2^{-n_q}.
\end{align}

Since the support window parameter $L_{\epsilon}$ increases as $\epsilon$ decreases (see~\cref{eq:tailwx,eq:tailwp}), for fixed $n_q$, the probability of having a successful transfer protocol, $P^{CD}(\epsilon)$, decreases with decreasing the protocol error. For fixed $P^{CD}$, ~\cref{eq:prob_failure} implies that the number of necessary qubits scales with the error as $n_q \propto \log_2 \left(L_\e\right)$. When the qumode's wavefunction decrease exponentially fast with increasing $\abs{x}$ and $\epsilon \propto e^{-c L_{\epsilon}}$ (which is a good approximation for Fock states, as the numerical calculations presented in \cref{fig:pgood}(b) and in Ref.\cite{macridin_pra_2021} show),  $n_q \propto \log_2 \left[\ln\left(\epsilon^{-1}\right)\right]$.

It is useful to investigate  the transfer of the Fock states  $\{\phi_n(x)\}_n$, since
our discretization method requires  the qumode to be truncated in the Fock states basis. In~\cref{fig:telecvdv} we illustrate the transfer of the Fock states of order $n=0$, $n=31$ and $n=62$ to a DV device with $n_q=7$ qubits.
As can be seen from
~\cref{fig:telecvdv}(a), for a fixed $\e$, the support interval parameter $L_{\epsilon}$ increases with increasing $n$. Consequently, the range of
$p_{meas}$ for high-fidelity transfer  is decreasing with increasing $n$, see ~\cref{fig:telecvdv}(b) and (c).  In ~\cref{fig:pgood} we investigate the probability of achieving high fidelity transfer.
The dependence of $L_{\epsilon}$ on the Fock state order $n$ is illustrated in~\cref{fig:pgood}(a),
while its dependence on the error $\epsilon$ is illustrated in ~\cref{fig:pgood}(b).
The behavior of $P^{CD}(\epsilon,n)$
as a function of  $n$ for fixed $\epsilon$ and as a function of $\epsilon$ for fixed $n$ is shown in ~\cref{fig:pgood}(c) and~\cref{fig:pgood}(d), respectively.

The transfer fidelity and probability of success monotonically decrease with increasing $n$.
Consequently, the transfer fidelity and probability of success for a qumode  with a cutoff $N_b$
are always better than those corresponding to the transfer of the Fock state of order $N_b$.
In other words, for a qumode with a cutoff $N_b$, a successful transfer protocol is guaranteed with the resources necessary for the transfer of the Fock state of order $N_b$.

As can be inferred from ~\cref{fig:pgood}(c), even when the number of qubits of the DV device is small, ({\em{i.e.}} $n_q=4,5,6$), the transfer protocol can be implemented  with significant probability, (\emph{e.g.} $P^{CD}(\epsilon=10^{-4}) > 0.1$), for CV states with boson cutoff $N_b < 20$. Presumably, this will make the experimental implementation of the transfer  protocol feasible with present or near-future technology.  On the other hand, for a near-deterministic protocol, a  high success probability is desired. In this case, the number of required qubits is  of the order of $20$. For example, for $P^{CD}=0.99$ ($P^{CD}=0.999$) and qumode states with a cutoff $N_b=100$, the necessary number of qubits $n_q > \approx 21$ ($n_q > \approx 28$) when the required precision is $\O(10^{-7})$.

The number of qubits needed for near-deterministic transfer protocol is larger than the number of qubits required for accurately representing the qumode on qubits. For instance, in the previous paragraph we found that the transfer of a Fock state with $n=100$  requires  $\approx 21$ qubits for a probability of success $P^{CD} \approx 0.99$. However, this state can be represented with an accuracy of $\O(10^{-25})$~\cite{macridin_pra_2021} on just $8$ qubits, meaning that an ancillary register of $\approx 13$ qubits was used to increase the transfer success probability.
In~\cref{sec:discard_qubits}, we will show how to discard the ancillary register after the transfer protocol is complete.

\subsection{Qumode transfer from DV device to CV device}
\label{sec:tele_dvcv}

\begin{figure}[tb]
    \begin{center}
        \includegraphics*[width=5in]{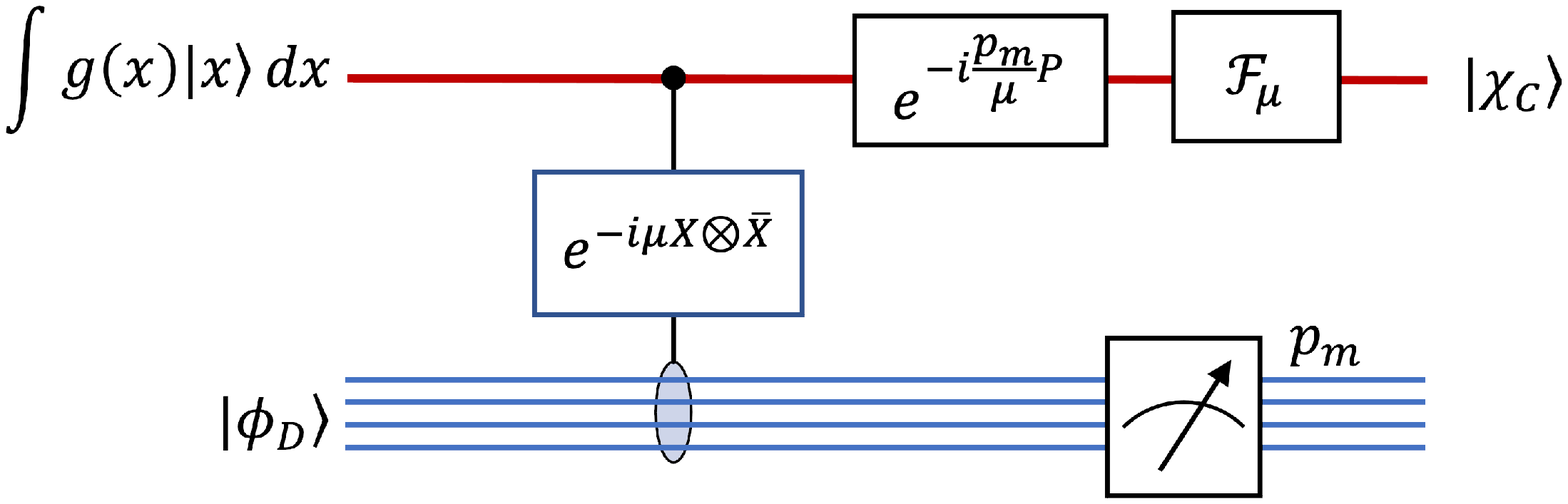}
        \caption{DV-CV transfer protocol, described in~\cref{sec:tele_dvcv}. The DV state
        $\ket{\phi_D}$ is transferred into the CV state $\ket{\chi_C}$.
        }
        \label{fig:diagdvcv}
    \end{center}
\end{figure}

The DV-CV transfer protocol, diagrammatically presented in~\cref{fig:diagdvcv}, consists of the following steps:
\begin{enumerate}
\item The CV state is prepared into
\begin{align}
\label{eq:cvini}
\int g(x) \ket{x}_C dx,
\end{align}
\noindent with $\int \abs{g(x)}^2 dx=1$. The joint DV-CV initial wavefunction reads
\begin{align}
\label{eq:cvdv_state2}
\ket{\chi_{DC}}= \sqrt{\Delta_x}\int  \sum_{j=0}^{N_x-1} g(x) \phi(x_j)\ket{x}_C\ket{j}_D dx.
\end{align}

The protocol success depends on the initial state of the CV device defined by the wavefunction $g(x)$. At the end of this section, we will discuss the choice of $g(x)$ and provide two examples.

\item The entangling operator
$e^{-i \mu X \otimes \X}$ is applied,
\begin{align}
\label{eq:cvdv_state}
 e^{-i \mu X \otimes \X} \ket{\chi_{DC}}= \sqrt{\Delta_x}\int g(x)  \sum_{j=0}^{N_x-1} \phi(x_j) e^{-i \mu x x_j}\ket{x}_C\ket{j}_D dx.
\end{align}

\item The DV system in measured in the discrete momentum basis.
Let's denote the measured value $p_m$. According to~\cref{eq:grid_p}, $p_m=(m-\frac{N_x-1}{2})\Delta_p$, with  $m\in\{0,...,N_x-1\}$. The CV state after the measurement is
\begin{align}
\label{eq:chi0}
\ket{\chi_{C0}}
&=\sqrt{\frac{\Delta_p}{Pr(p_m)}}\int
g(x)\hphi_{aper}(\mu x +p_m)\ket{x}_C dx,
\end{align}
\noindent where
\begin{align}
\label{eq:hphi_aper}
\sqrt{\Delta_p}\hphi_{aper}(\mu x +p_m)= \frac{1}{\sqrt{N_x}} \sum_{j=0}^{N_x-1} \sqrt{\Delta_x} \phi(x_j) e^{-i x_j \left( \mu x +p_m\right)},
\end{align}
\noindent and $Pr(p_m)$ is the probability to measure $p_m$,
\begin{align}
\label{eq:probdv}
Pr(p_m)=\Delta_p\int \abs{g(x)\hphi_{aper}(\mu x +p_m)}^2 dx.
\end{align}
The function $\hphi_{aper}(p)$ is anti-periodic
since $2L\sqrt{\mu}x_j =2 \pi \left(j-\frac{N_x}{2}+\frac{1}{2}\right)$ and $N_x=2^{n_q}$ is an even number.  Employing~\cref{eq:sh_dftphi}, we have
\begin{align}
\label{eq:period}
\hphi_{aper}(p)&=-\hphi_{aper}(p+2L\sqrt{\mu}), \\
\label{eq:perinside}
\hphi_{aper}(p)&=\hphi(p) ~\text{ when } ~ p \in \left[-L\sqrt{\mu}, L\sqrt{\mu}\right].
\end{align}

\item   The operator  $e^{-i \frac{p_m}{\mu} P}$ is applied to the CV system
\begin{align}
\label{eq:cvstate_1}
\ket{\chi_{C1}} =e^{-i \frac{p_m}{\mu} P}\ket{\chi_{C0}}&= \sqrt{\frac{\Delta_p}{Pr(p_m)}}  \int
g(x)\hphi_{aper}(\mu x +p_m)\ket{x+\frac{p_m}{\mu}}_C dx.
\end{align}

\item The continuous Fourier transform $\F_{\mu}$, defined as
\begin{align}
\label{eq:mcft}
\F_{\mu}=\sqrt{\frac{\mu}{2 \pi}} \int dx \int dy e^{i \mu xy} \ket{x}\bra{y},
\end{align}
\noindent that can be implemented
using phase shift and squeezing operations,  is applied to the CV system. The CV state becomes
\begin{align}
\label{eq:cvstate_2}
\T^{DC}(\ket{\phi_D})\equiv\ket{\chi_C}=\F_{\mu} \ket{\chi_{C1}}=  \int \xi(x) \ket{x}_C dx,
\end{align}
\noindent where
\begin{align}
\label{eq:xi}
\xi(x)&= \sqrt{\frac{\Delta_p}{Pr(p_m)}} \frac{1}{\sqrt{2 \pi \mu}} \int  g(\frac{k-p_m}{\mu})\hphi_{aper}(k) e^{i kx} dk.
\end{align}

\end{enumerate}

By employing the anti-periodicity property of $\hphi_{aper}(p)$, it can be shown that (see~\cref{app:atele_dvcv})
\begin{align}
\label{eq:xiform}
\xi(x) e^{-i x p_m}&= \frac{1}{\sqrt{Pr(p_m)}} \sqrt{\frac{\Delta_p}{N_x}} \sum_{j=-\infty}^{\infty}\phi(x_j)e^{-i x_j p_m}\hg\left[\mu \left(x_j-x\right)\right].
\end{align}
\noindent Here
\begin{align}
\hg(t)= \frac{1}{\sqrt{2 \pi}} \int  g(k) e^{-i k t} dk,
\end{align}
\noindent is the Fourier transform of $g(x)$. The probability to measure $p_m$
can be written as (see~\cref{app:atele_dvcv})
\begin{align}
\label{eq:probdvcv}
Pr(p_m)=\frac{\Delta_p}{N_x \mu} \sum_{i,j=-\infty}^{\infty}\phi^{*}(x_i)\phi(x_j)  e^{- i \left(x_i-x_j\right) p_m} \int  \hg^{*}(z+\mu x_i) \hg(z+\mu x_j) dz.
\end{align}

The protocol probability of success, defined as the probability of having a measurement outcome such that the fidelity is larger than $1-\epsilon$, is given by
\begin{align}
\label{eq:probDC_eps}
P^{DC}(\epsilon)=\sum_{m=0}^{N_x-1} Pr(p_m)\bigg|_{F_C>1-\epsilon}.
\end{align}
\noindent where $F_C$ is defined by~\cref{eq:teleFDC}.

By inspecting~\cref{eq:xiform}, it can be seen that $\xi(x) e^{-i x p_m}$ is, up to a normalization factor,
the convolution of
the set $\{\phi(x_j)e^{-i x_j p_m}\}_j$ with the function $\hg(\mu x)$. The next goal is to find appropriate
choices of $\hg(\mu x)$ such that $\xi(x) \approx \phi(x)$. We present two examples below.

\subsubsection{Rectangular initial CV state.}
\label{sec:recg}

\begin{figure}[tb]
    \begin{center}
        \includegraphics*[width=5in]{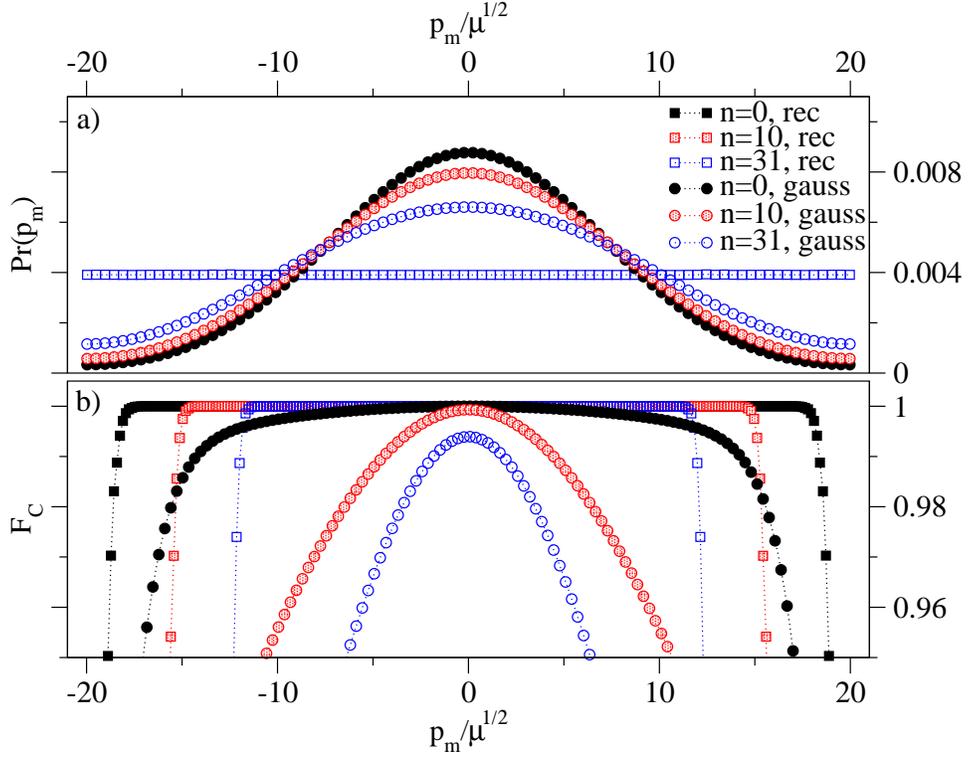}
        \caption{Transfer of $n=0$ (full symbols), $n=10$ (shaded symbols) and $n=31$ (open symbols) Fock states from a  $n_q=8$ qubit device to a CV device initially prepared with rectangular (rectangle symbols) and  Gaussian (circle symbols) wavefunctions. The Gaussian wavefunction has $\sigma=0.5 \frac{L}{\sqrt{\mu}}$ (see~\cref{eq:cvinigauss}). The dotted lines are for visual guidance.
       a) Probability to measure $p_m$. While for a rectangular CV initial state $Pr(p_m)$ is constant
       (see~\cref{eq:prdvrec}), it has a Gaussian shape for a Gaussian CV initial state, with a width
      that increases with increasing $n$. b) Transfer fidelity (\cref{eq:teleFDC}) versus $p_m$. For rectangular CV initial state the fidelity $F_C \ge 1-\O(\epsilon)$ when $\abs{p_m} \le L\sqrt{\mu}-L_{\epsilon}\sqrt{\mu}$. Compared to the rectangular case, for a Gaussian
       CV initial state the fidelity is smaller and decreases faster with increasing $p_m$ and $n$. }
        \label{fig:teledvcv}
    \end{center}
\end{figure}

If the Fourier transform of  $\phi(x) e^{-i x p_m}$ had support on the finite interval $p\in\left[-L\sqrt{\mu}, L\sqrt{\mu}\right]$
and $\hg(\mu x)$ were proportional to the {\em{sinc }} function $u(x)$ defined by~\cref{eq:sincu},  the Nyquist-Shannon theorem and~\cref{eq:xiform} would imply $\xi(x)=\phi(x)$. Therefore, our first choice of $g(x)$   is  the rectangular function
\begin{align}
\label{eq:grec}
g(x)=\left\{
  \begin{array}{ll}
  \frac{\mu^{\frac{1}{4}}}{\sqrt{2L}} &~~\text{for}~~~~ x \in \left[-\frac{L}{\sqrt{\mu}}, \frac{L}{\sqrt{\mu}}\right]\\
  0 &~~\text{for}~~~~ |x|>\frac{L}{\sqrt{\mu}}
  \end{array}
 \right. ,
\end{align}
\noindent because for this choice we have (see~\cref{eq:ftu} in~\cref{app:sinc})
\begin{align}
\label{eq:fthg}
\hg(\mu x)&=\frac{1}{\sqrt{\Delta_p}}u(x).
\end{align}
\noindent The orthogonality property of the {\em{sinc }} functions described by~\cref{eq:ortho} (\cref{app:sinc}), together with \cref{eq:probdvcv}, yields a probability to measure $p_m$ that is independent of $p_m$,
\begin{align}
\label{eq:prdvrec}
Pr(p_m)=\frac{\Delta_p}{N_x} \sum_{i=-\infty}^{\infty} \abs{\phi(x_i)}^2 \frac{\Delta_x}{\Delta_p} = \frac{1}{N_x}.
\end{align}

The Fourier transform of $\phi(x)e^{-ixp_m}$ is $\hphi(p+p_m)$. Since the $\epsilon$-support interval
of $\hphi(p)$ is $\left[-L_{\epsilon}\sqrt{\mu}, L_{\epsilon}\sqrt{\mu}\right]$, $\hphi(p+p_m)$ has negligible
({\em{i.e.}} $\O(\epsilon)$) support outside the interval $\left[-L\sqrt{\mu}, L\sqrt{\mu}\right]$, as long as
\begin{align}
\label{eq:pmvalid}
\abs{p_m} \le L\sqrt{\mu}-L_{\epsilon}\sqrt{\mu}.
\end{align}

In this case the DV state can be transferred with $O(\epsilon)$ precision to the CV register,
{\em{i.e.}}
\begin{align}
\label{eq:xi3}
\xi(x)&=  \phi(x)+\O(\epsilon),~\text{ when }~ \abs{p_m} \le L\sqrt{\mu}-L_{\epsilon}\sqrt{\mu}.
\end{align}

\Cref{eq:xi3,eq:probDC_eps} imply
\begin{align}
\label{eq:probDC_tele}
P^{DC}(\epsilon) \approx P^{DC}\left[(O(\epsilon)\right]=\frac{L-L_{\epsilon}}{L}=\frac{\sqrt{N_x}-L_{\epsilon}\sqrt{2/\pi}}{\sqrt{N_x}}.
\end{align}

For illustration, in~\cref{fig:teledvcv} we show (rectangle symbols) the probability $Pr(p_m)$
and the fidelity $F_C$ versus $p_m$ for the transfer of the Fock states with $n=0$, $n=10$ and $n=31$
from a $n_q=8$ qubits device to a CV device.

Note that, apart from the small term $\frac{\Delta_p}{2}$  and the fact that $p_m$ is discrete, \cref{eq:pmvalid} is similar to \cref{eq:pvalid} which gives the condition for high-fidelity CV-DV transfer protocol. Up to the small $\frac{1}{N_x}$ term, we also have $P^{DC}(\epsilon) \approx P^{CD}(\epsilon)$, as can be seen by comparing~\cref{eq:probCD_tele,eq:probDC_tele}. Practically, the dependence of DV-CV transfer protocol on the number of qubits and accuracy is the same as the corresponding dependence of CV-DV transfer protocol discussed in~\cref{sec:tele_cvdv} and illustrated in~\cref{fig:pgood} for the Fock states.

\subsubsection{Gaussian initial CV state.}
\label{sec:gaussg}

\begin{figure}[tb]
    \begin{center}
        \includegraphics*[width=5in]{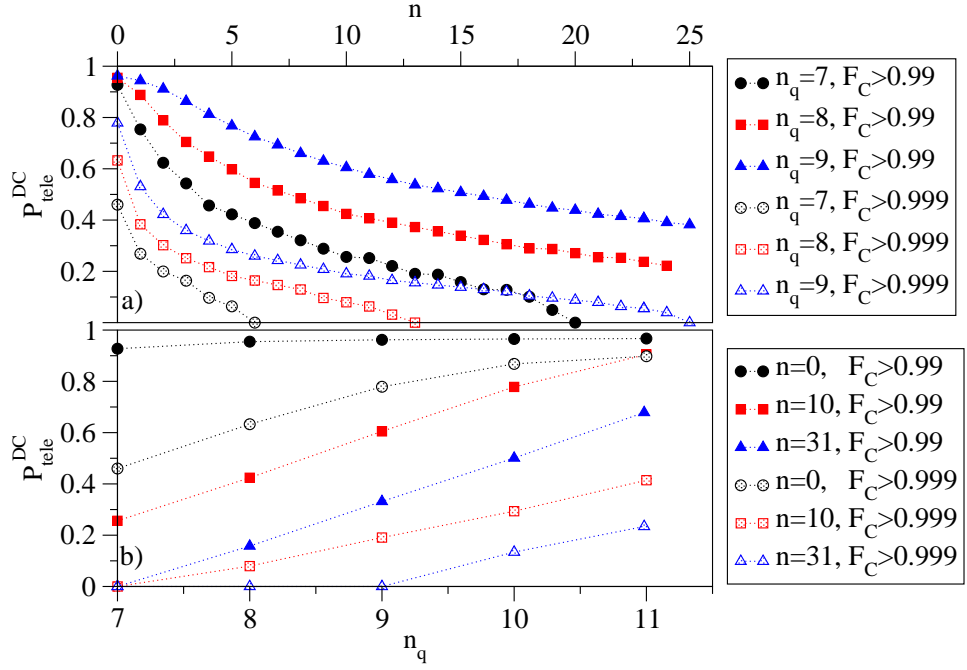}
        \caption{DV-CV transfer of Fock states when the initial CV states is a Gaussian with
        $\sigma=0.5 \frac{L}{\sqrt{\mu}}$.
        a) Probability $P^{DC}(\epsilon)$ (see~\cref{eq:probDC_eps}) versus $n$  for   $\epsilon=0.01$ (full symbols)
        and $\epsilon=0.001$ (shaded symbols)  when  $n_q=7,8$ and $9$. b) $P^{DC}(\epsilon)$ versus the number of qubits $n_{q}$, for the transfer of Fock states with $n=0,10$, and $31$ when $\epsilon=0.01$
        and $\epsilon=0.001$. $P^{DC}$ decreases as $n$ increases and increases as $n_q$ increases.
        }
        \label{fig:teledvcvP}
    \end{center}
\end{figure}

A rectangular initial state of the CV device ensures a high-fidelity  transfer protocol and a probability of success that approaches one exponentially fast as the number of qubits increases, similar to the CV-DV transfer protocol. However, preparing rectangular CV states might be challenging in practice, since a rectangular state is non-Gaussian. Here we show that the DV-CV transfer protocol works and can be brought to the near-deterministic regime for alternative initial CV wavefunctions, which can be easily prepared in practice, but at the cost of increasing the number of required qubits in the DV device. Namely we address the
DV-CV transfer when the initial CV state is a Gaussian function,
\begin{align}
\label{eq:cvinigauss}
g(x)&=\pi^{-\frac{1}{4}}\frac{1}{\sqrt{\sigma}}e^{-\frac{x^2}{2\sigma^2}}
\end{align}
\noindent with  variance $\sigma^2$.

For this choice of $g(x)$, \cref{eq:xiform} yields
\begin{align}
\label{eq:xi5}
\xi(x)e^{-i x  p_m}&=   \pi^{-\frac{1}{4}} \sqrt{\Delta_p\sigma} \sum_{j=-\infty}^{\infty}
\phi(x_j)e^{-i x_j  p_m} e^{-\frac{\mu^2 \sigma^2}{2}  \left(x-x_j \right)^2 }.
\end{align}

By inspecting \cref{eq:xi5}, we expect that
\begin{align}
\sigma \lessapprox\frac{1}{\mu\Delta_x}=\frac{L}{\sqrt{\mu}}
\end{align}
is required for a smooth convolution. On the other hand, a value of $\sigma$ that is too small will average out the variation of $\phi(x)e^{-i x p_m}$ along the grid points.
It is expected that as the variation of $\phi(x)$ and the value of $p_m$ increase, the protocol fidelity will decrease. This has been confirmed by numerical calculations. We  also have found numerically that $\sigma \in \left[0.5 \frac{L}{\sqrt{\mu}},0.6\frac{L}{\sqrt{\mu}}\right]$ yields the best protocol fidelity (not shown).

In \cref{fig:teledvcv}(a) and (b), we show (circles) the probability to measure $p_m$ and, respectively, the fidelity for the transfer of Fock states with $n=0$, $n=10$, and $n=31$ from an 8-qubit device to a CV device initially prepared in a Gaussian state with $\sigma=0.5\frac{L}{\sqrt{\mu}}$. The probability to measure $p_m$ has a Gaussian shape with a width that increases as $n$ increases. Compared to the rectangular initial CV state, the fidelity is smaller and decreases faster with increasing $\abs{p_m}$ and $n$.

The  transfer probability of success  is shown in \cref{fig:teledvcvP} for Fock states. $P^{DC}(\epsilon)$ for  fixed $\epsilon$ decreases with increasing $n$ and increases with increasing the number of qubits in the DV register.

Similar to the rectangular case, the accuracy and success probability can be increased by increasing $n_q$. However, for the same  level of accuracy, the number of qubits required is greater for the Gaussian case than for the rectangular case. We have not thoroughly investigated the dependence of the transfer fidelity and success probability on the number of qubits for Gaussian initial CV states, because a Gaussian initial CV state is not the only practical choice for DV-CV transfer protocol, and probably not the best one either. In a future study, we plan to investigate DV-CV transfer protocol for various initial states, such as variational available states or states consisting of a sum of displaced Gaussians.

\section{Ancillary qubits for near-deterministic transfer protocol}
\label{sec:ancillas}

As discussed in \cref{sec:tele_cvdv}, a high success probability and high fidelity  CV-DV transfer protocol requires a number of qubits significantly larger than the one necessary for an accurate discrete representation of the qumode. After the transfer, many coefficients of the discrete qumode state in the basis $\{\ket{j}_D\}_j$ with $j \in \{0,...,N_x-1\}$ are negligible. In \cref{sec:discard_qubits}, we show how to down-size the DV register to the minimum number of qubits required for the discrete representation of the qumode with the desired accuracy.

Similarly, for a high success probability and high fidelity DV-CV transfer protocol, the DV register should have a number of qubits significantly larger than the one necessary for the representation of the qumode to be transferred. In \cref{sec:add_qubits}, we show how to, in order to increase the success probability of the transfer protocol, add ancillary qubits to the DV register.

\subsection{Qubit discard after CV-DV transfer}
\label{sec:discard_qubits}

\begin{figure}[tb]
    \begin{center}
        \includegraphics*[width=5in]{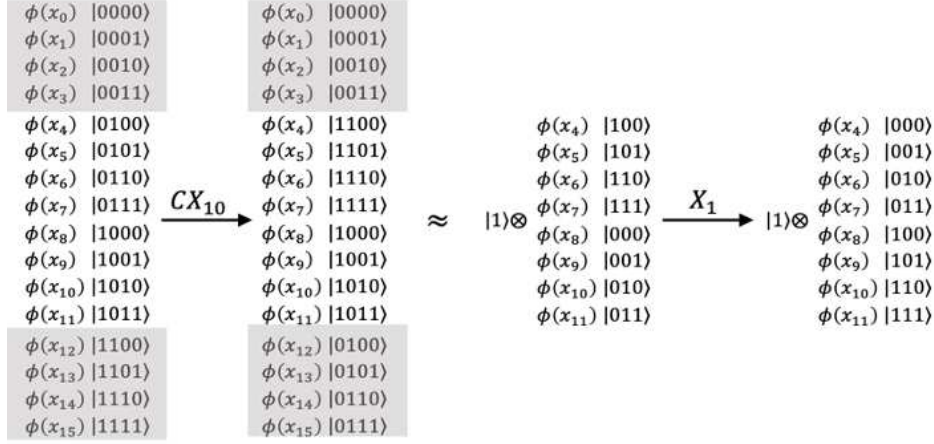}
        \caption{One qubit discard. The coefficients of the basis vectors $\{\ket{j}_D\}$ sown in the shaded region are negligible ($\approx\O(\epsilon)$). First,  a $CX$ gate is applied to the first two qubits (qubit $0$ and the control qubit $1$). Second, an $X$ gate is applied to the qubit $1$.
        As a result, all basis vectors with nonzero coefficients will have the qubit $0$ in the state $\ket{1}$. The $0$ qubit is unentangled and can be discarded. The remaining
        state is described by~\cref{eq:discard2}.}
        \label{fig:discard}
    \end{center}
\end{figure}

In order to achieve high-fidelity, near-deterministic CV-DV transfer, a DV register with a large number of qubits needs to be used. However, not all qubits are necessary to represent the qumode after the transfer. Here, we present a method for discarding unnecessary qubits. We will begin with the procedure for discarding one qubit.

As described in~\cref{sec:tele_cvdv},  after a successful high-fidelity
transfer protocol  the DV state is
\begin{align}
\label{eq:ancphiD}
\ket{\phi_D}=\sqrt{\Delta_x}\sum_{j=0}^{N_{x}-1} \phi(x_j)\ket{j}_D +\O(\epsilon),
\end{align}
\noindent with $x_j=\left(j-\frac{N_x-1}{2}\right)\Delta_x$ and  $\Delta_x=\sqrt{\frac{2 \pi}{N_x \mu}}$.
The goal of this procedure is to obtain the
 state
 \begin{align}
 \label{eq:phiprime}
\ket{\phi^{\prime}_D}=\sqrt{\Delta^{\prime}_x}\sum_{j=0}^{N^{\prime}_{x}-1} \phi(x^{\prime}_j)\ket{j}_D+\O(\epsilon),
\end{align}
\noindent on a DV device with $n^{\prime}_{q}=n_q-1$ qubits, where  $N^{\prime}_{x}=N_{x}/2$,
and $x^{\prime}_j=\left(j-\frac{N^{\prime}_x-1}{2}\right)\Delta^{\prime}_x$ with $\Delta^{\prime}_x=\sqrt{2}\Delta_x$.

A number of qubits larger than the one required for the
qumode discrete representation implies that the number of the discretization points $N_x$ is
large enough such that $\frac{L}{\sqrt{2}}\ge L_{\epsilon}$, with $L_{\epsilon}$ defined by~\cref{eq:tailwx,eq:tailwp} and $L$ defined by~\cref{eq:LNx}.
The coefficients $\phi(x_j) = \O(\epsilon)$ for $j\in \{0,...,\frac{1}{4}N_x-1\}$ and $j\in \{\frac{3}{4}N_x,...,N_x-1\}$,
because, for these values of $j$, $x_j$ is outside the $\epsilon$-support window of function $\phi(x)$,
$\left[-\frac{L_{\epsilon}}{\sqrt{\mu}},\frac{L_{\epsilon}}{\sqrt{\mu}}\right]$.
In our encoding, as defined by~\cref{eq:xjqub}, the qubits defining the basis states  are counted from left to right, {\em{i.e.}} $\ket{j}_D=\ket{j_0,j_1,...,j_{n_q-1}}$.
The first part of the procedure, as illustrated in~\cref{fig:discard}, consists in applying a $CX$ gate
to the qubits $1$ and $0$ (with $1$ being the control qubit), followed by an $X$ gate to the qubit $1$,
\begin{align}
\label{eq:discard1}
&\sqrt{\Delta_x}\sum_{j=0}^{N_x-1} \phi(x_j)\ket{j_0,j_1,...,j_{n_q-1}}
\xrightarrow[]{~CX_{10}} \sqrt{\Delta_x}\sum_{j=0}^{N_x-1}\phi(x_j)\ket{j_0 \oplus j_1,j_1,...,j_{n_q-1}} \\ \nonumber
& \xrightarrow[]{~~X_1~~} \sqrt{\Delta_x} \sum_{j=0}^{N_x-1} \phi(x_j)\ket{j_0 \oplus j_1,j_1 \oplus 1,...,j_{n_q-1}}
=\ket{1}\otimes \sqrt{\Delta_x}\sum_{j=\frac{N_x}{4}}^{\frac{3N_x}{4}-1} \phi(x_j)\ket{j_1 \oplus 1,...,j_{n_q-1}}+\O(\epsilon)\\ \nonumber
&=\ket{1}\otimes \sqrt{\Delta_x}\sum_{j=0}^{N^{\prime}_x-1} \phi(\tilde{x}_{j})\ket{j_0,j_1,...,j_{n_q-2}}+\O(\epsilon),
\end{align}
\noindent where $\tilde{x}_{j}=\left(j-\frac{N^{\prime}_x-1}{2}\right)\Delta_x=x^{\prime}_j/\sqrt{2}$,
and $\oplus$ denotes {\em{modulo}} $2$ summation.
After these two transformations, the qubit $0$ becomes unentangled and is discarded.

\begin{figure}[tb]
    \begin{center}
        \includegraphics*[width=5in]{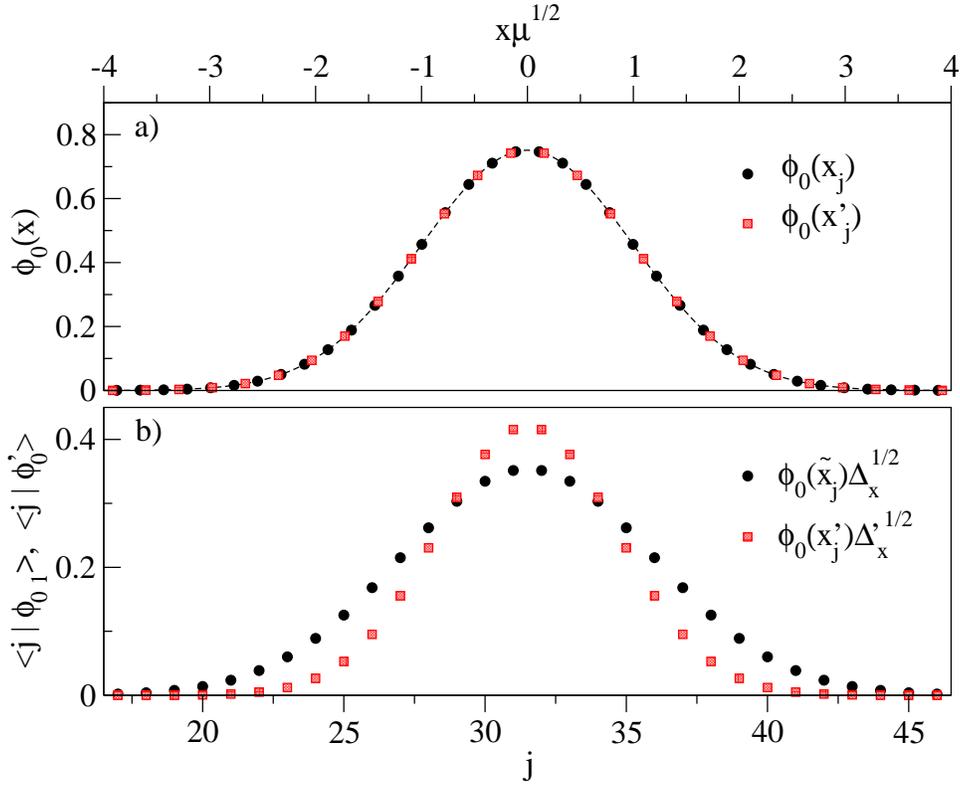}
        \caption{Hermit-Gaussian (Fock) wavefunction, $\phi_0(x)$, of order $n=0$. a) The black circles
        are sampled on a grid with the discretization distance $\Delta_x=\sqrt{\frac{2 \pi}{N_x}}$,
        where $N_x=128$, corresponding to $n_q=7$ qubits and $\mu=1$. The red squares are sampled on a grid with $\Delta^{\prime}_x=\sqrt{\frac{2 \pi}{N^{\prime}_x}}$ where $N^{\prime}_x=64$, corresponding to $n_q=6$ qubits and $\mu=1$.
        b) The value of the discretized qumode coefficients in the $n_q=6$ qubits basis $\{\ket{j}_D\}_{j \in \{0,...,63\}}$, before (black circles) and after (red squares)  the squeezing operation described by~\cref{eq:squeezeop} is applied.
        }

        \label{fig:squeeze}
    \end{center}
\end{figure}

After discarding the qubit, the DV state on $n_q-1$ qubits is
\begin{align}
\label{eq:discard2}
\ket{\phi_1}=\sqrt{\Delta_x}\sum_{j=0}^{N^{\prime}_x-1} \phi(\tilde{x}_{j})\ket{j}_D+\O(\epsilon).
\end{align}
However, this is not exactly the state we target, since the sampling points $\{\tilde{x}_{j}\}$ are on a grid
with the discretization interval $\Delta_x$, as illustrated with black-circle symbols in~\cref{fig:squeeze} for the $n=0$ Fock state. We want the sampling points for the target state
to be on a grid with the discretization interval $\Delta^{\prime}_x=\sqrt{2}\Delta_x$, illustrated
with red-square symbols in~\cref{fig:squeeze}.

The second part of the procedure consists in applying a squeezing gate
with the squeeze factor $r=\ln 2/2$. According to~\cref{eq:squeezephi} the state becomes
\begin{align}
\label{eq:squeezeop}
\S(\frac{1}{2}\ln 2)\ket{\phi_1}=
\sqrt{\Delta^{\prime}_x}\sum_{j=0}^{N^{\prime}_x-1} \phi(x^{\prime}_{j})\ket{j}_{D}+\O(\epsilon)
=\ket{\phi^{\prime}_D}+\O(\epsilon),
\end{align}
\noindent which, up to $\O(\epsilon)$ error, is just the qumode representation on $n_q-1$ qubits, as
described by~\cref{eq:phiprime}.

Note that, even before applying the squeezing operation, the qumode representation on the reduced qubit register described by~\cref{eq:discard2} is valid. However, it corresponds to a discretization for mass $\mu^{\prime}$-bosons, where $\mu^{\prime}=2 \mu$. In this representation, the discrete position and momentum operators should be defined as in~\cref{eq:Xd,eq:Pd}, but with $\mu^{\prime}$ replacing $\mu$. It is important to note that the $\mu$-boson and $\mu^{\prime}$-boson number distributions of the qumode are different. The representation with the lowest number of bosons cutoff is more accurate. A  more detailed discussion about the relation between the boson mass and the representation accuracy is presented in~\cite{macridin_pra_2021}.

The one-qubit discarding procedure described above can be repeated to discard more qubits.
The  number of qubits that can be discarded is equal to the maximum
integer $r$ that satisfies $\frac{L}{\sqrt{2^r}} \ge L_{\epsilon}$.

\subsection{Qubit padding before DV-CV transfer}
\label{sec:add_qubits}

The success probability of DV-CV transfer protocol increases with increasing size of the DV register. The procedure to add a qubit to the DV register consists of the same steps as the qubit discarding procedure presented in~\cref{sec:discard_qubits}, but in reverse order.

The $n_q$-qubit initial DV state is
\begin{align}
\label{eq:addq_ini}
\ket{\phi_D}=\sqrt{\Delta_x}\sum_{j=0}^{N_{x}-1} \phi(x_j)\ket{j}_D.
\end{align}
\noindent The $n_q+1$-qubit target  DV state is
 \begin{align}
 \label{eq:addq_target}
\ket{\phi^{\prime}_D}=\sqrt{\Delta^{\prime}_x}\sum_{j=0}^{N^{\prime}_{x}-1} \phi(x^{\prime}_j)\ket{j}_D+\O(\epsilon),
\end{align}
\noindent with $N^{\prime}_{x}=2N_x$,
$\Delta^{\prime}_x=\frac{\Delta_x}{\sqrt{2}}$, and $x^{\prime}_{j}=\left(j-\frac{N^{\prime}_x-1}{2}\right)\Delta^{\prime}_x$.

The first step of the padding procedure is  squeezing
with the squeeze factor $r=-\ln2/2$. According to~\cref{eq:squeezephi} the state becomes
\begin{align}
\label{eq:squeezeop1}
\ket{\phi_1}\equiv\S(-\frac{1}{2}\ln 2)\ket{\phi_D}=\sqrt{\frac{\Delta_x}{\sqrt{2}}}\sum_{j=0}^{N_x-1} \phi(\frac{x_{j}}{\sqrt{2}})\ket{j}_{D}+\O(\epsilon)=\sqrt{\Delta^{\prime}_x}\sum_{j=\frac{N^{\prime}_x}{4}}^{\frac{3N^{\prime}_x}{4}-1} \phi(x^{\prime}_{j})\ket{j}_{D}+\O(\epsilon).
\end{align}

Next, a qubit prepared in the state $\ket{1}$ is added to the left of the register, {\em{i.e.}}, $\ket{\phi_1} \longrightarrow \ket{1}\otimes\ket{\phi_1}$. According to the encoding convention defined by~\cref{eq:xjqub}, this new qubit will be in position $0$. Next, the steps shown in~\cref{fig:discard} are followed in reverse order, {\em{i.e.}}, an $X_1$ gate is applied to the qubit in position $1$, followed by a $CX_{10}$ gate applied to the qubits in positions $1$ and $0$.
This procedure yields the target state $\ket{\phi^{\prime}_D}$ described by~\cref{eq:addq_target}, up to an error given by the weight of $\phi(x)$ outside the interval $\left[-\frac{N_x\Delta_x}{2},\frac{N_x\Delta_x}{2}\right]$.

In order to increase the transfer protocol success probability to the desired value, the procedure described above can be repeated to add more qubits.

\section{Conclusions}
\label{sec:conclusions}

Qumodes are bosonic quantum states that encode information in the continuous basis formed by the eigenvectors of the quadrature operators. We introduce a discrete representation of the qumodes on the finite Hilbert space of DV devices, along with the implementation of the quadrature operators
and the implementation of a universal set of CV gates on DV devices.
We construct the discrete qumode representation by employing the Nyquist-Shannon expansion theorem, which is applicable to qumode wavefunctions that have negligible weight at large arguments.
The errors associated with this representation decrease exponentially with increasing the size of the finite Hilbert space when the qumode can be truncated in the boson number basis.

We present two protocols for transferring qumodes between CV and DV devices. The first protocol transfers a CV qumode to its discrete representation on a DV device. The protocol has high fidelity when the measurement outcome is confined to a specific interval. The probability of achieving high-fidelity transfer approaches one exponentially as the number of qubits in the DV register increases.

The second  protocol transfers a discrete DV qumode to a CV device. The fidelity of the protocol depends on the measurement outcome. If the initial CV device is prepared with a rectangular wavefunction, the dependence of the transfer fidelity and success probability on the number of DV qubits is practically the same as that of the CV-DV transfer protocol. For instance, in this case, the success probability approaches one exponentially as the number of qubits in the DV device increases.
However, we find that even with alternative initial CV states, which may be easier to prepare experimentally, the DV-CV transfer protocol can  be implemented with high fidelity and high success probability, albeit requiring more DV qubits.

The transfer protocols can be driven  to the near-deterministic regime by increasing the number of DV qubits. This can be achieved by using ancillary registers that can be discarded after the protocol is completed. We introduce procedures for discarding qubits after CV-DV transfer and for adding qubits before DV-CV transfer. These procedures consist of single-qubit gates, CNOT gates, and  squeezing operations.

Alternative methods for transferring qumodes between CV and DV devices may also be possible. For instance, a protocol for transferring Fock states below the cutoff between the CV and DV devices, along with a unitary transformation from the Fock basis to the discrete quadrature basis on the DV device, will be effective if the practical implementation of these protocols is feasible. To the best of our knowledge, there are no practical proposals in the literature for the implementation of such protocols. The number of gates necessary for the transformation between Fock and quadrature bases on a DV device scales as $\mathcal{O}(4^{n_q})$ (unless more efficient approximate unitaries can be found via variational methods). However, the transfer of the Fock states between the CV and DV devices appears even more challenging to implement. For example, the Fock states can be transferred using a SWAP algorithm or by a measurement-based protocol similar to the one used here, where instead of the quadrature variables $X$ and $P$, the number operator and the phase operator are used as conjugate variables~\cite{pegg_barnett_PRA_1989}. Apparently, the most difficult problem is the implementation of the phase operator~\cite{milburn_PRA_1999, Clausen_PRA_2000, Um_NC_2016} and the transformation from the number basis to phase states basis on CV devices.

The work presented in this paper demonstrates the potential of hybrid CV-DV quantum hardware for  processing  CV-encoded information, opening up new research directions for hybrid CV-DV systems and creating opportunities for developing integrated quantum technology. We envision a wide range of applications for this study. For example, CV-encoded data from optical or cavity sensors can be transferred to qubit QPUs and analyzed with QML methods that could be challenging to implement on CV devices. Non-Gaussian states can be transferred from a DV device to a CV device, and non-Gaussian gates can be realized by teleporting qubit gates implemented on DV devices to CV devices by developing  protocols similar with the ones described in~\cite{Bartlett_PRL_2003,Weedbrook_RMP84_2012}.
This would provide an efficient alternative to preparing CV states and gates directly, which typically requires nontrivial optimal pulse control~\cite{Krastanov2015}. Hybrid CV-DV cluster states can be employed for quantum computation.
The quantum tomography of CV states can be reduced  to an equivalent qubit system tomography problem,
by transferring the CV states to DV devices.
These are a few examples illustrating how CV-DV hybrid quantum hardware, with the proposed transfer protocols, could make quantum information processing more efficient.
We believe that this work will enable the development of a new class of quantum algorithms using CV-DV hybrid hardware in various fields such as quantum computing, quantum networking, quantum sensing, quantum tomography, and quantum machine learning.

\section{Acknowledgements}

This material is based upon work supported by the U.S. Department of Energy, Office of Science, National Quantum Information Science Research Centers, Superconducting Quantum Materials and Systems Center (SQMS) under contract number DE-AC02-07CH11359. A.C.Y.L. is partially supported by the DOE/HEP QuantISED program grant "HEP Machine Learning and Optimization Go Quantum", identification number 0000240323.
This manuscript has been authored by Fermi Research Alliance, LLC under Contract No. DE-AC02-07CH11359 with the U.S. Department of Energy, Office of Science, Office of High Energy Physics.

\appendix

\section{Nyquist-Shannon expansion  with shifted grid sampling}
\label{app:NShlf1}

Here we show that the Nyquist-Shannon expansion for band-limited function remains valid if the
sampling grid is shifted by an arbitrary amount.

Let $f(x)$ be a band-limited function, {\em i.e.} $\hf(p)=0$  for $|p| > L$,
where $\hf(p)$ is the Fourier transform of $f(x)$  and $L$ is a positive real number.
The Nyquist-Shannon theorem~\cite{Shannon_1949} implies
\begin{align}
\label{eq:NSf}
f(x)=\sum_{i=-\infty}^{\infty} f(x_i) u(x-x_i),
\end{align}
\noindent where $x_i=i \Delta_x$. The  grid interval $\Delta_x$ and the function $u(x)$  are defined by~\cref{eq:sincu,eq:delta_x}, respectively, with $L\equiv L_\e$ and $\mu=1$.

Now, let's define the function $h(x)$ as follows:
\begin{align}
\label{eq:fshift}
h(x):=f(x+\delta),
\end{align}
\noindent where $\delta$ is an arbitrary real number. Its  Fourier transform,
\begin{align}
\hh(p)=\frac{1}{\sqrt{2 \pi}}\int dx f(x+\delta) e^{-ipx}=e^{ip\delta}\hf(p),
\end{align}
\noindent  has support on the interval $\left[-L,L\right]$, similar to $\hf(p)$. The Nyquist-Shannon theorem  applied to $h(x)$ implies that
\begin{align}
\label{eq:NSg}
h(x)=\sum_{i=-\infty}^{\infty} h(x_i) u(x-x_i).
\end{align}
\noindent \Cref{eq:NSg} is equivalent to
\begin{align}
\label{eq:NSfdelta}
f(x+\delta)=\sum_{i=-\infty}^{\infty} f(x_i+\delta) u(x-x_i).
\end{align}
\noindent By changing the variable $x+\delta \longrightarrow x$ in  \cref{eq:NSfdelta} we get
\begin{align}
\label{eq:NSgridsh}
f(x)=\sum_{i=-\infty}^{\infty} f(x_i+\delta) u(x-x_i-\delta).
\end{align}
The band-limited function $f(x)$ can be expressed as an infinite sum with the sum terms proportional
to the function sampled on the grid points $\{i \Delta_x+\delta\}_{i\in \mathbb{Z}}$. The parameter $\delta$ is
a real arbitrary number.

\section{Some properties of the {\em{sinc}} functions}
\label{app:sinc}

Here we present two properties of the {\em{sinc}} functions employed
in the Nyquist-Shannon expansion of  band-limited functions, that are useful to our study.

The Fourier transform of a {\em{sinc}} function is a rectangular function.  Thus,
\begin{align}
\label{eq:ftu}
\int_{-\infty}^{\infty}  u(x-x_j) e^{-i p x} d x=e^{-i p x_j} R(p) \Delta_x ,
\end{align}
\noindent where $R(p)$
\begin{align}
 R(p)=\left\{
  \begin{array}{ll}
  1 &~~\text{for}~~~~ p \in \left[-L\sqrt{\mu}, L\sqrt{\mu}\right]\\
  0 &~~\text{for}~~~~ |p|>L\sqrt{\mu}
  \end{array}
 \right. ,
 \end{align}
\noindent as can be directly checked.

The {\em sinc} functions $u(x-x_j)$ defined by~\cref{eq:sincu}  obey the orthogonality relation
\begin{align}
\label{eq:ortho}
\int u(x-x_j) u(x-x_l) dx = \Delta_x \delta_{jl}.
\end{align}
\noindent \Cref{eq:ortho} ca be obtained by employing the Parseval-Plancherel theorem and~\cref{eq:ftu}.

\section{Shifted finite Fourier transform}
\label{app:DFT_sh}

As described by~\cref{eq:phix,eq:phip}, the Nyquist-Shannon theorem implies that, when a boson cutoff can be imposed, the  wavefunction $\phi(x)$ and its Fourier transform $\hphi(p)$ can be expressed as finite sums. The sum terms are proportional to $\phi(x)$ sampled at the grid points $\{x_j+\delta_x\}_{j \in \{0,...,N_x-1\}}$ and to $\hphi(p)$ sampled at the grid points $\{p_m+\delta_p\}_{m \in \{0,...,N_x-1\}}$, respectively. The Nyquist-Shannon expansion is valid for any grid shift parameters $-0.5<\delta_x, \delta_p \le 0.5$. The function sampled at the position grid points and the function's Fourier transform sampled at the momentum grid points are connected via shifted discrete Fourier transforms, as described by~\cref{eq:sh_dftphi,eq:sh_dfthpi}.

In order to represent the qumode on a DV device, in~\cref{sec:fhsp} we construct a finite Hilbert space
of dimension $N_x$ by defining a basis $\{\ket{j}\}_{j \in \{0,...,N_x-1\}}$ and the action of the discrete
quadrature operators in this basis. On this finite Hilbert space,  the shifted Fourier transform is  defined as follows:
\begin{align}
\label{eq:sft}
\tF_{\delta_x,\delta_p}=\frac{1}{\sqrt{N_x}} \sum_{k,j=0}^{N_x-1}    e^{i \frac{2 \pi}{N_x} \left(j-\frac{N_x-1}{2}+\delta_x\right) \left(k-\frac{N_x-1}{2}+\delta_p\right)}  \ket{j}\bra{k}.
\end{align}
The explicit implementation of the shifted Fourier transform on qubits is given in~\cref{sec:sftqubits}.

When both shift parameters are zero, {\em{i.e.}} $\delta_x=\delta_p=0$, we obtained the
centered discrete Fourier transform,
\begin{align}
\label{eq:dft}
\tF \equiv \tF_{0,0}=\frac{1}{\sqrt{N_x}} \sum_{k,j=0}^{N_x-1}
e^{i \frac{2 \pi}{N_x} \left(j-\frac{N_x-1}{2}\right) \left(k-\frac{N_x-1}{2}\right)}  \ket{j}\bra{k}.
\end{align}
\noindent The centered Fourier transform is used to define the discrete momentum operator in~\cref{eq:Pd}. Note that the discrete momentum operator defined in this way obeys  the parity symmetry-related equations
\begin{align}
\mu\F^{-1}\X \F = -\P\\
\frac{1}{\mu}\F^{-1}\P \F = \X\\
\frac{1}{\mu}\F\P \F^{-1} = -\X,
\end{align}
\noindent  similar to the equations satisfied by the continuous operator  $P$~\cite{Weedbrook_RMP84_2012}.

Employing~\cref{eq:phix} we can write
\begin{align}
\label{eq:phix1x2}
\phi(x_i+\delta_{x1}\Delta_{x}) = \sum_{j=0}^{N_x-1} \phi(x_j+\delta_{x2} \Delta_{x}) u\left[x_i-x_j+\left(\delta_{x1}- \delta_{x2}\right)\Delta_{x}\right]+\O(\epsilon),
\end{align}
The  sets $\{\phi(x_i+\delta_{x1})\}_i$ and $\{\phi(x_j+\delta_{x2})\}_j$  are connected
by the operator $\tF_{\delta_{x1},\delta_p} \tF^{-1}_{\delta_{x2},\delta_p}$. This implies
that
\begin{align}
\label{eq:top}
T_{\delta_{x1},\delta_{x2}}\equiv\tF_{\delta_{x1},\delta_p} \tF^{-1}_{\delta_{x2},\delta_p}=\sum_{i,j=0}^{N_x-1} u\left[x_i-x_j+\left(\delta_{x1}- \delta_{x2}\right)\Delta_{x}\right] \ket{i}\bra{j} + \O(\epsilon)
\end{align}
\noindent when acting on the subspace defined by the cutoff $N_b$.

In particular, the operator  $T_{\delta_{x},0}$ acting on a discrete qumode yields
\begin{align}
\label{eq:Topqum}
T_{\delta_{x},0} \left[\sqrt{\Delta_x} \sum_{j=0}^{N_x-1} \phi(x_j) \ket{j}\right]=\sqrt{\Delta_x} \sum_{j=0}^{N_x-1}
\phi(x_j+\delta_x\Delta_x) \ket{j}+ \O(\epsilon),
\end{align}
\noindent and provides access to the wavefunction values on shifted grid points
by employing measurements in the DV device computational basis.

\subsection{Implementation of shifted discrete Fourier transform on qubits}
\label{sec:sftqubits}

The shifted discrete Fourier transform reduces to the implementation of the standard Quantum Fourier transform sandwiched between single-qubit ${\cal{R}}^z$ rotations. In order to show this, we write~\cref{eq:sft} as
\begin{align}
\label{eq:sft1}
\tF_{\delta_x,\delta_p}=e^{i A(\delta_x,\delta_y)} \left[ \sum_{k=0}^{N_x-1} e^{-i k B(\delta_p)}\ket{k}\bra{k}\right] QFT \left[ \sum_{j=0}^{N_x-1} e^{-i j B(\delta_x)}\ket{j}\bra{j}\right]
\end{align}
\noindent with
\begin{align}
\label{eq:aphase}
 A(\delta_x,\delta_p)&=\frac{2 \pi}{N_x} \left(\frac{N_x-1}{2}-\delta_x\right)\left(\frac{N_x-1}{2}-\delta_p\right),\\
 \label{eq:bphase}
 B(\delta)&=\frac{2 \pi}{N_x} \left(\frac{N_x-1}{2}-\delta\right),
\end{align}
\noindent and
\begin{align}
\label{eq:qft}
QFT=\frac{1}{\sqrt{N_x}}\sum_{k,j=0}^{N_x-1}  e^{i \frac{2 \pi}{N_x} jk}\ket{j}\bra{k}.
\end{align}

The first term in~\cref{eq:sft1} is a phase factor.
The third term in~\cref{eq:sft1} is the standard Quantum Fourier Transform, and its implementation is described in~\cite{nielsen2002quantum}. The main computational cost for its implementation
is given by the $n_q(n_q-1)/2$ two-qubit CNOT gates.

The second  and the fourth terms are diagonal operators. Their implementation requires only single-qubit rotation gates. For example, considering the encoding of our basis vectors on qubits, described by~\cref{eq:jqub}, we can  write
\begin{align}
\label{eq:bterm}
\sum_{k=0}^{N_x-1} e^{-i k B(\delta)}\ket{k}\bra{k}&=
\sum_{k_0,...,k_{n_q-1}=0,1} e^{-i \sum_{q=0}^{n_q-1} k_q 2^{n_q-1-q} B(\delta)}\ket{k_0,...,k_{n_q-1}}\bra{k_0,...,k_{n_q-1}} \\ \nonumber
&= \prod_{q=0}^{n_q-1} \left(\ket{0}\bra{0}_q+e^{-i 2^{n_q-1-q} B(\delta)}\ket{1}\bra{1}_q\right)
=e^{-i \frac{B(\delta)}{2}\left(N_x-1\right)}\prod_{q=0}^{n_q-1} {\cal{R}}^z_q\left[- 2^{n_q-1-q} \frac{B(\delta)}{2}\right],
\end{align}
\noindent where the ${\cal{R}}^z_q\left(\t \right)$ rotation acting on  qubit $q$ is
\begin{align}
\label{eq:rz}
{\cal{R}}^z_q\left(\t \right)=e^{-i \frac{\t}{2}\sigma^z_q}=e^{-i \frac{\t}{2}}\ket{0}\bra{0}+e^{i \frac{\t}{2}}\ket{1}\bra{1}.
\end{align}

\section{Implementation of CV gates on qubit devices}
\label{sec:gates}

The CV gates are mapped on DV devices by replacing the quadrature operators $X$ and $P$ with their discrete counterparts, $\X$ and $\P$, respectively. Here we present the explicit implementation of a set of gates which is sufficient for CV universal quantum computation.

By using~\cref{eq:Xdqub}, we get
\begin{align}
\label{eq:xdispl}
e^{-i \eta \X}=\prod_{q=0}^{n_q-1} R^z_{q}\left(- 2^{n_q-1-q} \Delta_x \eta \right),
\end{align}
\noindent with $R^z_{q}$ defined by~\cref{eq:rz}.

Analogously,
\begin{align}
\label{eq:x2gate}
e^{-i \eta \X^2}=e^{-i \eta \Delta_x^2 \frac{N_x^2-1}{12} } \prod_{p=0}^{n_q-1}\prod_{q=0}^{p-1} \gZZ_{pq} \left(\eta \nu_{pq} \right),
\end{align}
where
\begin{align}
	\gZZ_{pq}(\nu) = & \, e^{-i \nu \sigma^z_{p} \sigma^z_{q}}, \\
	\label{eq:nu}
	\nu_{pq} = & \, 2^{2n_q-3-p-q}  \Delta_x^2 .
\end{align}
The two-qubit gate  $\gZZ_{pq}$ acts on qubits $p$  and  $q$ and  can be decomposed into two CNOT gates and one $R^z$ gate \cite{Welch2014}. The gate $e^{-i \eta \X^2}$ consists of $n_q(n_q-1)$ CNOT gates.

The cubic-phase gate reduces to
\begin{align}
\label{eq:x3gate}
e^{-i \eta \X^3}=
\prod_{p=0}^{n_q-1}\prod_{q=0}^{p-1}\prod_{r=0}^{q-1}  \gZZZ_{pqr}(\eta \mu_{pqr})
\prod_{s=0}^{n_q-1} R^z_{s}(\eta \lambda_{s})
\end{align}
where
\begin{align}
	\gZZZ_{pqr}(\mu) = & \, e^{-i \mu \sigma^z_{p} \sigma^z_{q} \sigma^z_{r}}, \\
	\mu_{pqr} = & \, - 2^{3n_q - 4 - p -q - r}  \Delta_x^3, \\
	\lambda_{s} = & \,  -  2^{n_q - 3 - s} (  2^{2n_q - 2 - 2s} +   2^{2n_q} -  1 ) \Delta_x^3 .
\end{align}
The three-qubit gate $\gZZZ_{pqr}$ can be decomposed into four CNOT gates and one $R^z$ gate \cite{Welch2014}, and hence $e^{-i \eta \X^3}$ consists of $\frac{2}{3} n_q (n_q +1) (n_q +2)$ CNOT gates.

\Cref{eq:Pd} implies that  any gate $G(\P)$
function of the momentum operator $\P$, can be written as
$G(\mu\X)$ sandwiched between two centered quantum Fourier transform:
\begin{align}
\label{eq:opgate}
G(\P)=\tF G(\mu\X) \tF^{-1}.
\end{align}
\noindent In particular we can write
\begin{align}
\label{eq:p2gate}
e^{-i \eta \P^2}=\tF e^{-i \eta \mu^2 \X^2} \tF^{-1}.
\end{align}
\noindent The implementation of the gate $e^{-i \eta \P^2}$ reduces to the implementation of the
$e^{-i \eta \mu^2 \X^2}$ gate  described by~\cref{eq:x2gate},
and the implementation of the centered quantum Fourier gate described by ~\cref{eq:sft1} with $\delta_x=\delta_p=0$.

The CPHASE gate couples two different modes. On a DV device each mode is represented on a separate
$n_q$ qubit register. The CPHASE gate coupling the mode $i$ and the mode $j$
is implemented as
\begin{align}
\label{eq:cphase}
e^{-i \eta \X_i\otimes \X_j}=\prod_{p=0}^{n_q-1}\prod_{q=0}^{n_q-1}
	\gZZ_{pi;qj} \left(\eta \nu'_{pq} \right)
\end{align}
\noindent where
\begin{align}
	\gZZ_{pi;qj}(\nu) &=  \, e^{-i \nu \sigma^z_{pi} \sigma^z_{qj}} \\
	\label{eq:nuij}
	\nu'_{pq} &=2^{2n_q-4-p-q} \Delta_x^2.
\end{align}
 The $\gZZ_{pi;qj}$ gate acts on  qubit $p$ belonging to the qubit register allocated for mode $i$  and on qubit $q$ belonging to the qubit register allocated for mode $j$. This gate consists of $n_q^2$ $\gZZ$ gates or $2 n_q^2$ CNOT gates.

\section{Squeezing operator}
\label{sec:squeezing}

We pay particular attention to the implementation of the squeezing operator~\cite{gerry_knight_2004}
\begin{align}
\label{eq:sqop}
S(r)=e^{i \frac{r}{2}\left( X P+ P X \right)},
\end{align}
on a DV device. The discrete squeezing operator,
\begin{align}
\S(r) =e^{i \frac{r}{2}\left( \X \P+ \P \X \right)},
\end{align}
\noindent is used in the process of adding and discarding ancilla qubits to the DV device, as described in~\cref{sec:ancillas}.

The squeezing operator action on the quadrature operators is described by
\begin{align}
\label{eq:sqX}
S(r)^{\dagger} X S(r)& = X e^{-r}, \\
\label{eq:sqP}
S(r)^{\dagger} P S(r)&=P e^{r}.
\end{align}
\noindent \Cref{eq:sqX} implies
\begin{align}
S(r)^{\dagger} \ket{x} = e^{\frac{r}{2}}\ket{e^{r}x},
\end{align}
\noindent which is equivalent to
\begin{align}
\label{eq:sqchi}
\ket{\chi}=S(r)\ket{\phi} \implies \chi(x)=e^{\frac{r}{2}}\phi(x e^{r}).
\end{align}

Since squeezing can increase the number of bosons by a large amount, when mapping squeezing operations onto a DV device, it is always important to check whether  the number of discretization points $N_x$ is large enough to accurately represent squeezed states. Assuming we have chosen an $N_x$ large enough, the state $\ket{\chi}$ given by~\cref{eq:sqchi} will be represented on a DV device as follows
\begin{align}
\ket{\chi} =\sqrt{\Delta_j e^{r}}  \sum_{j=0}^{N_x-1}  \phi(x_j e^{r}) \ket{j}.
\end{align}
\noindent This implies that
\begin{align}
\S(r) \sqrt{\Delta_j} \sum_{j=0}^{N_x-1}  \phi(x_j) \ket{j}=\sqrt{\Delta_j e^{{r}}} \sum_{j=0}^{N_x-1}  \phi(x_j e^{r}) \ket{j}.
\end{align}

\subsection{Implementation of the squeezing operator}

\begin{figure}[tb]
	\begin{center}
		\includegraphics*[width=0.6\linewidth]{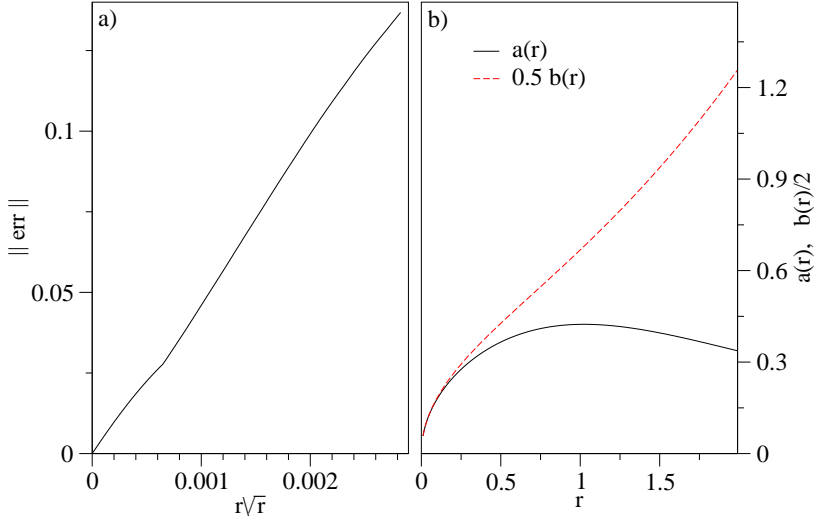}
		\caption{a) The error of the standard decomposition  of the squeezing operator, $\abs{\abs{err}}$ versus $r\sqrt{r}$,  calculated by imposing a boson occupation cutoff equal to $50$. The norm
       $\abs{\abs{err}} \equiv \max_{jk} \abs{err_{jk}}$, where $err_{jk}$ is the $\left(j,k\right)$ matrix element of the operator $err$ defined by \cref{eq:S_decomposition_appro}.
      b) Coefficients $a$ and $b$ defining the exact decomposition of the squeezing operator described by~\cref{eq:S_exact}, determined by solving numerically~\cref{eq:sq_a_and_b}.}
		\label{fig:squeeze_decompo}
	\end{center}
\end{figure}

While in CV devices the squeezing operation is implemented directly by using different experimental methods~\cite{slusher_PRL_1985,Shelby_PRL_1986,Wu_PRL_1986,SCHNABEL_PysRep_2017}, such as optical parametric down-conversion, for example~\cite{Wu_PRL_1986}, the implementation of the discrete squeezing operator requires decomposition into gates that can be implemented on DV devices.

Any gate which is a polynomial function of $X$ and $P$ can be obtained from
a universal set of gates, such as the one given by~\cref{eq:xdispl,eq:x2gate,eq:x3gate,eq:p2gate,eq:cphase},
by using the operator relation~\cite{Lloyd_PRL82_1999}
\begin{align}
\label{eq:CVuniv}
e^{-i \t A} e^{-i \t B} e^{i \t A} e^{i \t B}&=e^{ \t^2 \left[A,B\right]} + \O(\t^3)
\end{align}
In particular, since
\begin{align}
\left[X^2, P^2\right]=2i\left(XP+PX\right)
\end{align}
\noindent the squeezing operator can be written as
\begin{align}
\label{eq:S_decomposition_appro}
S(r)=e^{\frac{r}{4} \left[iX^2, -iP^2\right]}
=e^{-i \frac{\sqrt{r}}{2} X^2} e^{i \frac{\sqrt{r}}{2} P^2} e^{i \frac{\sqrt{r}}{2} X^2} e^{-i \frac{\sqrt{r}}{2} P^2} +err.
\end{align}
\noindent where  $err=\O(r\sqrt{r})$. For illustration, in~\cref{fig:squeeze_decompo}(a)
we plot $\abs{\abs{err}}$ versus $r\sqrt{r}$. The result has been obtained by employing numerical calculations.

However, there is a better way to implement the squeezing operator. As described in the second part of this section, we have found that $S(r)$
can be \emph{exactly} decomposed   into a product of exponentials of $X^2$ and $ P^2$.
That is, we have:
\begin{align}
\label{eq:S_exact}
	S(r)  =
	\begin{cases}
		e^{-i  \frac{a(r)}{2} X^2 } e^{i \frac{ a(r) }{2} P^2}
		e^{i \frac{b(r)}{2} X^2 } e^{- i\frac{b(r)}{2} P^2}
		e^{-i \frac{a(r)}{2} X^2 } e^{i\frac{a(r)}{2} P^2}
		& \text{\ for } r \ge 0,
		\\
		e^{-i  \frac{a(|r|)}{2} P^2} e^{i \frac{ a(|r|)}{2} X^2}
		e^{i \frac{b(|r|)}{2} P^2 } e^{- i\frac{b(|r|}{2} X^2}
		e^{-i \frac{a(|r|)}{2} P^2 } e^{i\frac{a(|r|)}{2}  X^2}
		& \text{\ for } r < 0.
	\end{cases}
\end{align}
\noindent where the
the real coefficients $a(r)$ and $b(r)$  satisfy the system of equations
\begin{align}
\label{eq:sq_a_and_b}
\begin{cases}
&\frac{a[a^2 \left(b^2+1\right)-2 a b+b^2+2]-b}{\mathrm{st}(a, b)}=0\\
&\frac{
r - \frac{
\left[a^4 \left(b^2+1\right)-2 a^3 b+a^2 \left(b^2+2\right)+b^2\right]
\cot^{-1}\left(\frac{a^4-2 \left(a^2+2\right) a b+4 a^2+\left(a^4+3 a^2+1\right) b^2+2}{\sqrt{\left[a^4-2 \left(a^2+2\right) a b+4 a^2+\left(a^4+3 a^2+1\right) b^2+2\right]^2-4}}\right)
}
{
\sqrt{\left[\left(a^2+2\right)^2-2  \left(a^2+2\right) a b+\left(a^4+3 a^2+1\right) b^2\right] \left[\left(a^2+2\right) a b+\left(a^2+4\right) a^2+\left(a^4+3 a^2+1\right) b^2-2 \right]}
}
}{\mathrm{st}(a, b)}=0
\end{cases}.
\end{align}
\noindent  We solved~\cref{eq:sq_a_and_b} numerically. The denominator $\mathrm{st}(a, b) = \sqrt{\left[a^4-2 \left(a^2+2\right) a b+4 a^2+\left(a^4+3 a^2+1\right) b^2+2\right]^2-4}$ was included to stabilize the root finding. The coefficients $a(r)$ and $b(r)$ are plotted in~\cref{fig:squeeze_decompo}(b).

Finally, to implement the discrete squeezing operator defined by~\cref{eq:dsqeeze}, we replace $X$ and $P$ in~\cref{eq:S_exact} with $\X$ and $\P$, respectively, and then use~\cref{eq:x2gate,eq:p2gate} to
implement the exponentials of $\X^2$ and $\P^2$.

We will end this section by sketching the derivation of \cref{eq:S_exact,eq:sq_a_and_b}. We start with the Baker-Campbell-Hausdorff formula~\cite{BCH_formula}:
\begin{align}
\label{eq:bch1}
e^{Z} = e^{A} e^{B},
\end{align}
\noindent where $Z = A + B + \frac{1}{2}[A,B] + \frac{1}{12}[A, [A, B]] - \frac{1}{12}[B, [A, B]] + \cdots$.
The $\cdots$ notation indicates terms proportional to all possible higher-order commutators of $A$ and $B$.
Note that the set formed by the operators
\begin{align}
	h = \frac{1}{4} [X^2, P^2],~~ e = \frac{i}{2} X^2 \text{\ and \ },~  f = - \frac{i}{2} P^2.
\end{align}
is closed under the commutation operation, {\em{i.e.}}
\begin{align}
[h, e] = 2 e,~~  [h, f] = -2f~ \text{\ and \ } [e, f] =h.
\label{eq:sl2r_commutation_relation}
\end{align}

The closure property described by \cref{eq:sl2r_commutation_relation} and the Baker-Campbell-Hausdorff formula imply that any product of the exponentials of $h$, $e$, and $f$ can be written as the exponential of a linear combination of $h$, $e$, and $f$. For example, we can write:
\begin{align}
   \label{eq:bch}
	e^{-a e} e^{-a f} e^{b e} e^{b f}  e^{-c e} e^{-c f} = e^{x(a,b,c) e + y(a,b,c) f + z(a,b,c) h},
\end{align}
\noindent where $a$, $b$, and $c$ are real numbers, and $x$, $y$, and $z$ are real functions to be determined. \Cref{eq:S_exact,eq:sq_a_and_b} for $r \ge 0$ are derived by solving $x(a,b,c)= y(a,b,c) =0$ and $z(a,b,c) = r$. To find the solution  we notice that the commutation relations given by \cref{eq:sl2r_commutation_relation} are the same as the commutation relations of the generators of the special linear group $\mathrm{SL}_{2}(R)$ \cite{Harish1952}. These generators have a simple $2 \times 2$ real matrix representation,
\begin{align}
\label{eq:sl2r}
h = \left(
\begin{array}{cc}
	1 & 0 \\
	0 & -1 \\
\end{array}
\right),~~
e = \left(
\begin{array}{cc}
	0 & 1 \\
	0 & 0 \\
\end{array}
\right),~~
f = \left(
\begin{array}{cc}
	0 & 0 \\
	1 & 0 \\
\end{array}
\right)
.
\end{align}
Using~\cref{eq:sl2r}  in ~\cref{eq:bch}, we obtain $a=c$, together with  the nonlinear equations~\cref{eq:sq_a_and_b}.
The $r < 0$ case  follows from the observation that the commutation relations described by~\cref{eq:sl2r_commutation_relation} are preserved by the transformation
\begin{align}
	(h, e, f) \longleftrightarrow (-h, f, e),
\end{align}
\noindent which implies  that the second line in~\cref{eq:S_exact} can be obtained by replacing
$(r, X^2, P^2)$ with $(-r, P^2, X^2)$ in the first line of~\cref{eq:S_exact}.

\section{Explicit calculation of the measurement probability and wavefunction of DV-CV transfer protocol}
\label{app:atele_dvcv}

The DV-CV transfer protocol described in~\cref{sec:tele_dvcv}  takes a DV state described by the~\cref{eq:dvstate}
to a CV state described by~\cref{eq:cvstate_2} with
\begin{align}
\label{eq:xiapp}
\xi(x)&= \sqrt{\frac{\Delta_p}{Pr(p_m)}} \frac{1}{\sqrt{2 \pi \mu}} \int  g(\frac{k-p_m}{\mu})\hphi_{aper}(k) e^{i kx} dk \\
\label{eq:xiapp1}
&= \sqrt{\frac{\Delta_p}{Pr(p_m)}} \sqrt{\frac{\mu}{2 \pi}} \int  \phi_{aper}(t) \hg(\mu t-\mu x) e^{-i \left(t-x \right) p_m} dt.
\end{align}
\noindent In~\cref{eq:xiapp1}
\begin{align}
\hg(t)= \frac{1}{\sqrt{2 \pi}} \int dk g(k) e^{-i k t},
\end{align}
\noindent and

\begin{align}
\label{eq:phiat}
\phi_{aper}(t)&=\frac{1}{\sqrt{2 \pi}} \int dz \hphi_{aper}(z) e^{itz}= \frac{1}{\sqrt{2 \pi}}
\sum_{j=-\infty}^{\infty}\int_{-L\sqrt{\mu}+2jL\sqrt{\mu}}^{-L\sqrt{\mu}+2(j+1)L\sqrt{\mu}}
\hphi_{aper}(z) e^{itz}
\\ \nonumber&=\frac{1}{\sqrt{2 \pi}}\sum_{j=-\infty}^{\infty}e^{2ijtL\sqrt{\mu}} (-1)^j \int_{-L\sqrt{\mu}}^{L\sqrt{\mu}}
\hphi(z) e^{itz} dz=\phi(t)\sum_{j=-\infty}^{\infty}e^{i j \left(2L\sqrt{\mu} t+\pi\right)} +err(N_x)
\\ \nonumber&=\frac{\pi}{L\sqrt{\mu}}\phi(t)\sum_{j=-\infty}^{\infty}\delta\left(t-\frac{\left(2j+1\right)\pi}{2L\sqrt{\mu}}\right)+err(N_x)\\ \nonumber
&=\Delta_x \phi(t)\sum_{j=-\infty}^{\infty}\delta\left[t-(j+\frac{1}{2})\Delta_x\right] +err(N_x).
\end{align}
In ~\cref{eq:phiat} we used the anti-periodicity of the $\hphi_{aper}(z)$ function described by~\cref{eq:period,eq:perinside},
the Poisson summation formula
\begin{align}
\label{eq:poissonaper}
\sum_{j=-\infty}^{\infty} e^{i  j (T x+\pi)}
=\frac{2 \pi}{T}\sum_{j=-\infty}^{\infty}\delta\left(x-\frac{ \pi \left(2j+1\right)}{T}\right),
\end{align}
\noindent and
\begin{align}
\frac{1}{\sqrt{2 \pi}}\int_{-L\sqrt{\mu}}^{L\sqrt{\mu}} \hphi(z) e^{itz} dz=\phi(t)+err(N_x),
\end{align}
\noindent where $err(N_x)$ is of the order of the wavefunction $\hphi(z)$ weight outside the interval
$\left[-L\sqrt{\mu}, L\sqrt{\mu}\right]$.

By employing~\cref{eq:phiat} in~\cref{eq:xiapp1} the teleported  wavefunction writes as
\begin{align}
\label{eq:xi1}
\xi(x)&= \frac{1}{\sqrt{Pr(p_m)}} \sqrt{\frac{\Delta_p}{N_x}} \sum_{j=-\infty}^{\infty}\phi(x_j)\hg\left[\mu \left(x_j-x\right)\right]e^{-i \left(x_j-x \right) p_m}+err(N_x).
\end{align}

Similarly, the probability to measure $p_m$ given by~\cref{eq:probdv} writes as
\begin{align}
Pr(p_m)=\Delta_p\int dz \abs{h(z)}^2,
\end{align}
\noindent where
\begin{align}
\label{eq:az}
h(z)&=\frac{1}{\sqrt{2 \pi}} \int \phi_{aper}(t)\hg(z+\mu t) e^{- i t p_m} dt\\ \nonumber
&=\frac{\Delta_x}{\sqrt{2 \pi}}\sum_{j=-\infty}^{\infty}\phi(x_j) \hg(z+\mu x_j)e^{- i x_j p_m}+err(N_x).
\end{align}

\Cref{eq:az} implies
\begin{align}
Pr(p_m)&=\frac{\Delta_p}{N_x \mu} \sum_{i,j=-\infty}^{\infty}\phi^{*}(x_i)\phi(x_j)  e^{- i \left(x_i-x_j\right) p_m} \int  \hg^{*}(z+\mu x_i) \hg(z+\mu x_j) dz+err(N_x).
\end{align}

\end{document}